# Multitask Bayesian Neural Networks for Multiparameter Protein Engineering

**Fabio Herrera-Rocha**[1], **David Medina-Ortiz**[1], **Desiree Wyrzykala**[1], **Tharun Srinivasan Sudha**[1], **and Mehdi D. Davari**[1*]

[1]Department of Bioorganic Chemistry, Leibniz-Institute of Plant Biochemistry, Weinberg 3, 06120 Halle (Saale), Germany

## Abstract

Simultaneously engineering multiple protein properties remains a major challenge. Existing machine learning-based pipelines for protein engineering often model properties separately, failing to capture their dependencies and trade-offs. Here, we systematically evaluate how Bayesian parameterization on Multitask Neural Networks can enable robust simultaneous protein engineering under scarce, noisy experimental data. We curated a comprehensive set of 27 multiparameter protein datasets. Then, we compared three algorithm architectures spanning low to full Bayesian parameterization across 16 sequence representations and dimensionality reduction (2,592 models). Bayesian Last Layer models delivered the strongest overall accuracy, generalization, and calibration, ranking as the top-performing model on 70% of benchmark datasets. Dimensionality reduction improved predictive performance by up to 42% and enhanced calibration up to 57% across architectures. Notably, simple One-Hot encoding achieved top performance on 25% of benchmark datasets, particularly with larger datasets. These results establish practical design principles for reliable and data-efficient multiparameter protein engineering.

## 1 Introduction

Engineering proteins for biotechnological and industrial applications often requires optimizing multiple properties (e.g., activity, stability, selectivity, etc.) simultaneously across a limited number of experimentally characterized protein variants [1, 2]. Finding functional enhanced variant is considerably complex due to dependencies and trade-offs between properties [3, 4]. To improve efficiency, increasingly protein engineering methods is supported by machine learning (ML) methods to accelerate directed evolution by increasing screening efficiency and reduce time and costs [3, 4].

The main purpose of ML approaches in protein engineering is to move most of the screening burden *in silico* [3, 4]. Typically, these algorithms focus on predicting a single property and necessitate the training of multiple ML algorithms when accounting for multiple properties together[5]. Solving this multiparameter optimization problem with limited experimental data represents one of the main challenges in protein engineering.

Multitask learning (MTL)—an underexplored alternative—can capture the commonalities between different variant properties in a single framework from the input amino acid sequences[6–8]. This method additionally enables the sharing of information between related tasks to enhance the overall performance of all tasks using common input features[8]. This is particularly useful when a limited number of samples is available for ML training[7].

*`mehdi.davari@ipb-halle.de`

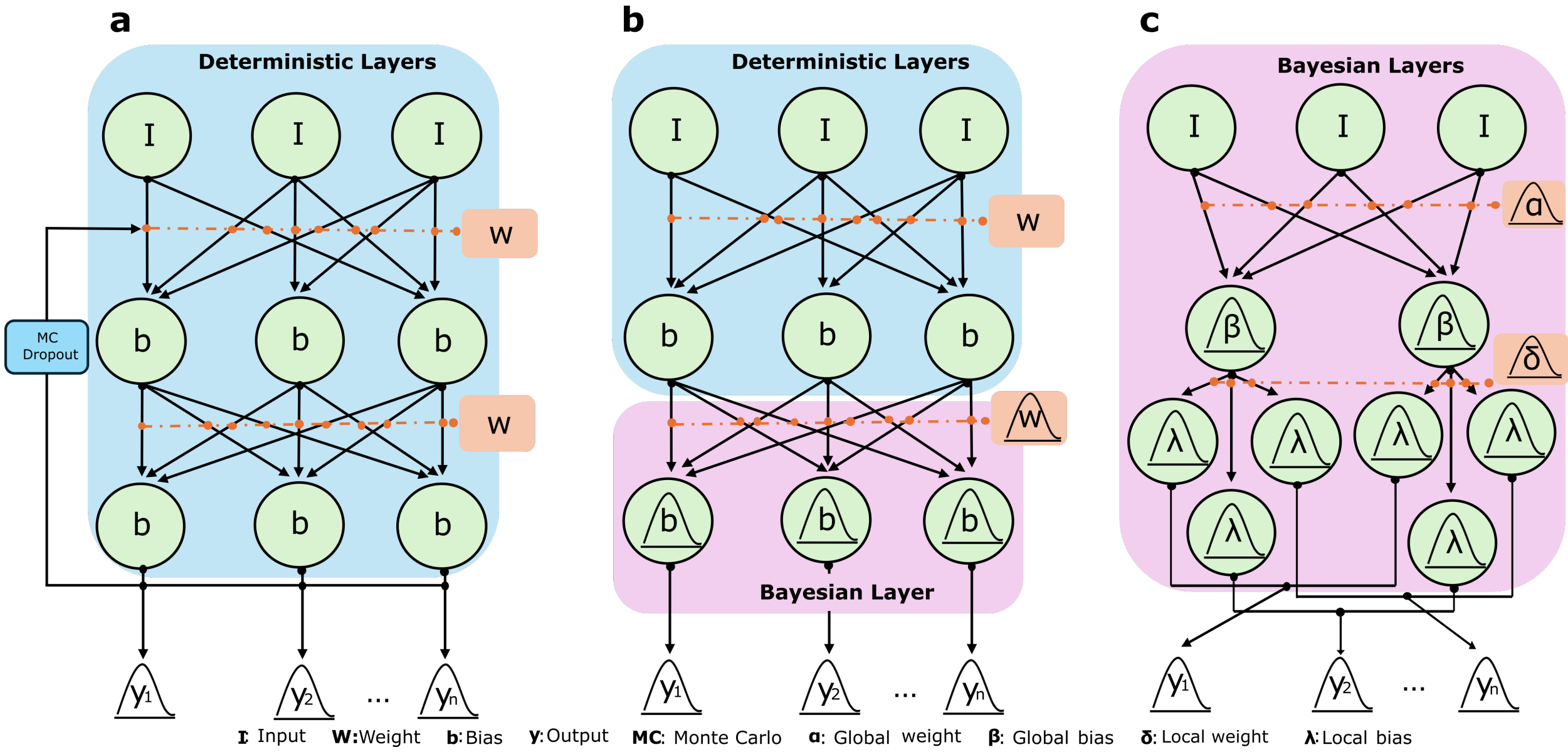

**Figure 1**: **Multitask Neural Network architectures with different degrees of Bayesian parameterization. (a)** Deterministic neural network with Monte Carlo dropout (MCD), where dropout is applied during inference and multiple stochastic forward passes ($n$ Monte Carlo samples) are used to approximate predictive uncertainty without explicit Bayesian parameter inference. **(b)** Bayesian Last Layer (BLL) neural network consisting of a deterministic feature extractor and a Bayesian output layer, in which posterior parameter sampling in the final layer is used to generate predictive distributions and estimate uncertainty. **(c)** Hierarchical Bayesian neural network (HBN) with shared global representations and task-specific hierarchical components for multitask prediction, where Bayesian inference is performed across all network levels to model parameter uncertainty and generate predictive distributions.

Bayesian Neural Networks (BNNs) offer a complementary solution to the challenges of multitask protein engineering by explicitly modeling uncertainty. Unlike conventional neural networks, BNNs treat model parameters as random variables rather than fixed point estimates [9]. This approach enables posterior predictive distributions and provides both predictions and associated uncertainty estimates [9–11]. This is particularly relevant to protein engineering, where experimental datasets are often scarce and noisy. Reliable uncertainty estimates can help distinguish confident predictions from poorly supported ones [11]. Thus, combining MTL with BNNs could simultaneously exploit shared information across protein properties while accounting for uncertainty arising from limited and noisy experimental data.

However, Bayesian MTL introduces an important challenge: determining the optimal degree of Bayesian parameterization within the neural network. Different degrees of Bayesian parameterization offer different trade-offs between model complexity, computational cost, predictive performance, and uncertainty estimation (Figure **1**). Monte Carlo dropout (MCD) [12] provides a low-cost approximation to Bayesian inference, whereas Bayesian Last Layer (BLL) [13, 14] restricts Bayesian inference to the output layer, and Hierarchical Bayesian Networks (HBN) [15] extend Bayesian inference across multiple hierarchical levels. Nevertheless, it remains unclear which level of Bayesian parameterization is most suitable under the limited and heterogeneous data regimes typical of multiparameter protein engineering.

This gap is further aggravated by the lack of standardized multiparameter protein datasets to compare these approaches. Here, we addressed both limitations by establishing the most comprehensive collection of standardized, ready-to-use multiparameter protein variant datasets following FAIR principles and using it to systematically benchmarked Bayesian MTL approaches. We evaluated deterministic, MCD, BLL, and HBN architectures under a unified validation framework, evaluating how Bayesian parameterization affects predictive performance, generalization, and uncertainty calibration across different dataset sizes and levels of task relatedness. By identifying when Bayesian integration provides a measurable advantage—and which degree of parameterization is most effective—we established practical design principles for reliable, data-efficient multiparameter protein engineering.

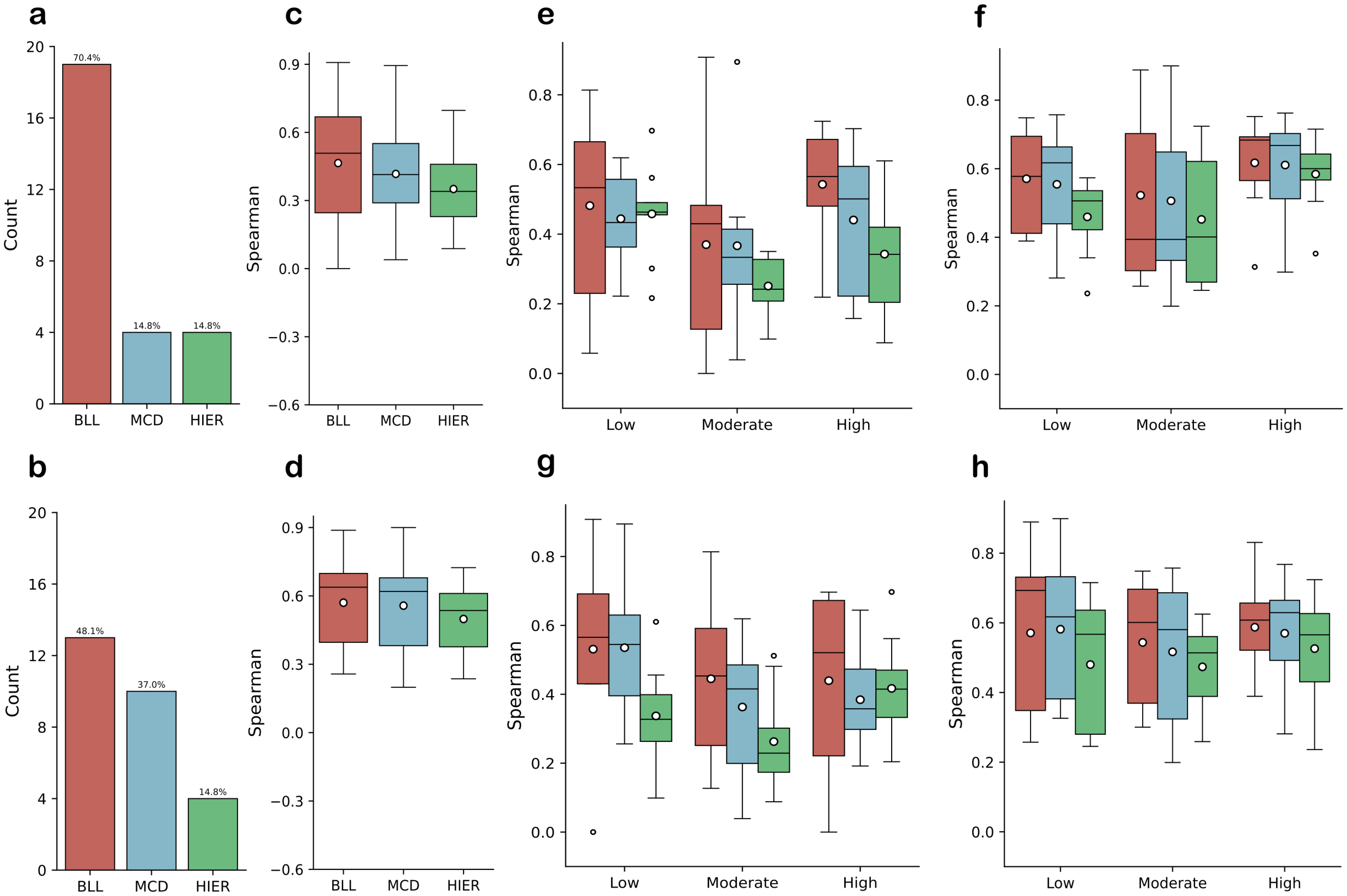

**Figure 2**: **Comparison of Multitask Bayesian Neural Network architectures for multiparameter protein engineering.** Performance of Bayesian Last Layer (BLL, red), Monte Carlo Dropout (MCD, blue), and Hierarchical Bayesian Neural Networks (HBN, green) across datasets. **a, b** Number of datasets where each architecture achieved the best model, without **(a)** and with **(b)** Supervised-PCA dimensionality reduction. (**c, d**) Distribution of Spearman correlation coefficients for the top model per architecture, without **(c)** and with **(d)** Supervised-PCA. (**e, f**) Performance stratified by dataset size—Low (<100 entries), Moderate (100–1000 entries), and High (>1000 entries)—without **(e)** and with **(f)** Supervised-PCA.(**g, h**) Performance stratified by label correlation—Low (Pearson <0.2), Moderate (0.2–0.5), and High (>0.5)—without **(g)** and with **(h)** Supervised-PCA. The white point in all the boxplots marks the mean Spearman correlation value.

## 2 Results

### 2.1 Benchmarking Multitask Bayesian Neural Network (MTBNNs) for multiparameter protein engineering

We benchmarked three BNN architectures with different levels of Bayesian parameterization—BLL, MCD, and HBN (Figure 1)—for multiparameter protein engineering tasks. MCD also serves as a baseline model because it is fundamentally a deterministic neural network that does not explicitly model Bayesian parameters [12]. Nevertheless, it provides predictive uncertainty estimates [12], making it a useful baseline for comparing uncertainty quantification and calibration performance against medium and fully Bayesian architectures. To evaluate our model performance, we compiled 27 benchmarking protein engineering datasets containing protein variants with two or more experimentally measured phenotypic properties. These datasets span a broad range of sizes, from 41 to more than 12,000 entries, and include between 2 and 16 simultaneously measured properties (Supplementary Table 1).

Protein sequences were represented using One-Hot encoding and embeddings derived from 15 pLM embeddings, resulting in 16 numerical representations per dataset. To mitigate data leakage [16], datasets were partitioned using a PCA-based similarity strategy using 41 global physicochemical properties of the amino acid sequence (see Methods section). Using the 27 datasets, 16 sequence representations, and 3 BNN architectures, we trained 1,296 models.

Because feature dimensionality can substantially affect ML models performance, all sequence representations were additionally compressed using supervised principal component analysis (SPCA) [17]. The training was repeated with SPCA-transformed feature space, resulting in 1,296 additional models and a total of 2,592 trained models. Model performance was evaluated using the Spearman correlation coefficient, which measures the ability to correctly rank protein variants rather than simply predict absolute fitness values[18].

We first assessed the frequency with which each BNN architecture achieved the best performance per datasets with the best sequence numerical representation. BLL was the top-performing architecture for 70% of the datasets (Figure **2**a), whereas MCD and HBN each accounted for approximately 15% of top-performing models. After SPCA dimensionality reduction, the proportion of datasets for which BLL achieved the best performance decreased to approximately 48%, whereas MCD increased to approximately 37%. The frequency of top-performing HBN models remained approximately constant (Figure **2**b).

Next, we compared the distribution of Spearman correlation coefficients across datasets. BLL showed the highest average performance, although differences between architectures were not statistically significant across architectures based on non-parametric global and pairwise comparisons (Kruskal-Wallis and Wilcoxon tests with Benjamini-Hochberg FDR correction;Figure **2**c). A similar pattern in average performance was observed after SPCA dimensionality reduction (Figure **2**d). A comparison of model performance before and after dimensionality reduction showed that all three architectures achieved significant improvements (Wilcoxon signed-rank test; adjusted $P < 0.001$ for MCD and HBN, and $P < 0.05$ for BLL), with performance increases of approximately 23%, 34%, and 42% for BLL, MCD, and HBN, respectively. These results indicate a consistent and beneficial effect of dimensionality reduction on predictive performance across all architectures.

### 2.2 Effect of dataset size and label correlation on MTBNNs performance

To examine how dataset size influences model performance, we grouped the benchmarking datasets into three size regimes: Low (<100 entries), Moderate (100–1,000 entries), and High (>1,000 entries), with 9 datasets in each category, approximately balanced across the full benchmark dataset pool. Evaluating performance across different data regimes is critical for MTBNNs applications in protein engineering, where experimentally characterized datasets are often limited in size [19–21].

Across all dataset sizes, BLL was the most frequent top-performing architecture, achieving the best performance in majority of datasets within each category (Supplementary Figure 1a). By contrast, HBN were the top-performing model only in a subset of Low-size datasets and did not outperform other architectures in Moderate- or High-size regimes.

Analysis of Spearman correlation coefficients showed that all the evaluated architectures achieved comparable average performance across data regimes, with no significant difference between them based on pairwise test with FDR correction (Figure **2**e). Notably, some BLL models in Low- and Moderate-size datasets reached Spearman correlation coefficients values comparable to, and in some cases exceeding, those observed in larger datasets, highlighting the capacity of this architecture to maintain strong predictive performance under limited-data conditions.

Following SPCA feature compression, BLL remained the most frequent top-performing architecture in Low- and High-size datasets, whereas MCD became the most frequent top architecture in the Moderate-size category (Supplementary Figure 1b). Although BLL retained the highest average performance, global differences between architectures were no longer statistically significant after dimensionality reduction based on FDR correction. SPCA significantly improved MCD models performance across dataset sizes (adjusted $P$ <0.05; Figure **2**f). The largest improvements were observed in HBN models in Moderate and High-data regimes, with 80% and 70% respectively, which showed both increased predictive accuracy and more stable performance.

We also investigated how the correlation between experimental labels affected model performance. Assessing label correlation is particularly relevant, as the strength of MTBNNs arise from their ability to exploit shared information across related outputs[22]. Datasets were therefore categorized according to the Pearson correlation coefficient between target labels as Low (<0.2), Moderate (0.2–0.5), or High (>0.5), resulting in 7, 8, and 12 datasets respectively.

MCD was the most frequent top-performing architecture in datasets with Low label correlation, accounting for 57% of top-performing models, whereas BLL was the top architecture for the remaining top-performing models in this category(Supplementary Figure 1c). BLL accounted for all top-performing models in Moderate and for the majority of top-performers in High label correlation datasets, whereas HBN achieved top performance only in a small number of High label correlation datasets.

Interestingly, performance analysis across correlation regimes showed that MCD achieved its highest average Spearman correlation coefficient in datasets with Low label correlation (Figure **2**g). Overall, performance tended to decrease with increasing label correlation from low to Moderate and High levels. HBN showed significantly lower (adjusted $P<0.05$) performance in Low- and Moderate-correlation datasets based on pairwise comparison with FDR correction, whereas no significant differences between architectures were observed for highly correlated labels. The highest-performing datasets were primarly concentrated in the upper quartile of BLL and MCD distributions within the Low-correlation category.

After SPCA dimensionality reduction, MCD further increased its prevalence as top performing architecture in Low-label correlation datasets, accounting for 71% of top-performing models(Supplementary Figure 1d). BLL remained the most frequent top-performing architecture for Moderate-correlation datasets, although with reduced prevalence. Minimal change was observed in the distribution of top-performing architectures in the High correlation regime (Figure **2**h), with the largest gains observed for High correlation datasets. Although BLL continued to show the highest average performance overall, differences between architectures were no longer statistically significant after dimensionality reduction.

### 2.3 Generalization capacity of the Bayesian neural network architectures

To assess the generalization capacity of the evaluated architectures, we quantified the generalization gap as the difference between the Spearman correlation coefficient obtained on the validation and test sets. This comparison provides an estimate of overfitting by evaluating the extent to which models maintain predictive performance on unseen data[23, 24]. Although some reduction in performance between validation and test sets is expected, large discrepancies indicate limited generalization capacity[24].

All evaluated architectures generally exhibited low overfitting scores (Figure **3**a), with no significant differences between them and an average overfitting score of 0.112. Across all architectures, overfitting tended to decrease with increasing dataset size, with particularly low scores observed in high-data regimes. In contrast, the highest overfitting scores were most pronounced in Low-data regimes (adjusted $P<0.01$; Figure **3**c). Similarly, only minor differences were observed between architectures across label-correlation categories (Figure **3**e). Overall, the overfitting trends broadly paralleled the general predictive performance of the models, as architectures with lower overfitting scores across multiple categories also showed superior predictive accuracy.

Under SPCA dimensionality reduction, only BLL models showed a significant increase in overall overfitting (adjusted $P<0.01$; Figure **3**b), whereas MCD and HBN architectures remained comparatively stable. The increase in overfitting observed for BLL was most pronounced in Low-data regimes ($P<0.05$; Figure **3**d) and was not detected across most label-correlation categories (Figure **3**f). In contrast, MCD and HBN models displayed minimal changes in overfitting across all dataset-size and label-correlation categories after dimensionality reduction.

The changes in overfitting after dimensionality reduction were broadly consistent with the overall performance trends observed for the different architectures. In particular, the increased overfitting of BLL models was accompanied by a reduction in the number of datasets for which BLL achieved top performance, whereas MCD architectures, which benefited most from dimensionality reduction, showed a corresponding increase in the frequency of top-performing models.

### 2.4 General performance of the sequence representations

To evaluate the effect of sequence representation, we compared the performance of all sequence representations across the three MTBNN architectures by analyzing both the distribution of Spearman correlation coefficients and the most frequent representations that produced the top-performing model across the 27 benchmarking datasets.

Across all architectures, a broad diversity of representations contributed top-performing models, with between 10 and 14 different embeddings reaching top performance before and after dimensionality reduction (Supplementary Figure 2). These results indicate that representation effectiveness is strongly dataset depen-dent. Nevertheless, several consistent trends emerged. In particular, One-Hot encoding consistently ranked among the best-performing representations despite its simplicity, often achieving performance comparable to large pLMs embeddings containing between 650 million and 5.7 billion parameters.

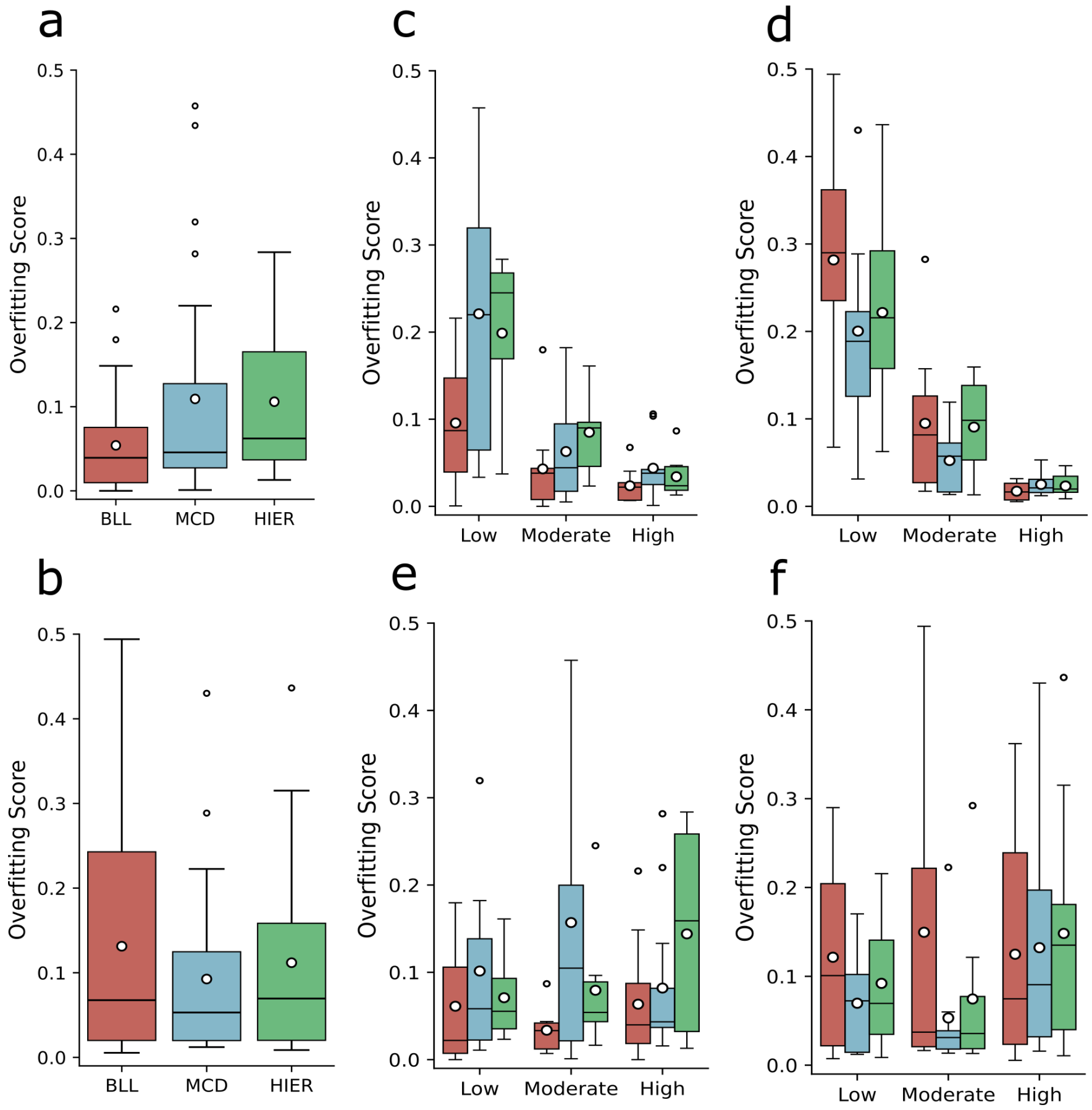

**Figure 3**: **Overfitting analysis of Multitask Bayesian Neural Network Architectures.** Overfitting of Bayesian Last Layer (BLL, red), Monte Carlo Dropout (MCD, blue), and Hierarchical Bayesian Networks (HBN, green) across benchmarking datasets. Overfitting was quantified as the difference between validation- and test-set Spearman correlation coefficients. **a, b** Distribution of the overfitting score for top models per architecture, without **(a)** and with **(b)** Supervised-PCA dimensionality reduction. (**c, d**) Overfitting score stratified by dataset size—Low (<100 entries), Moderate (100–1000 entries), and High (>1000 entries)—without **(c)** and with **(d)** Supervised-PCA. (**e, f**) Overfitting score stratified by label correlation—Low (Pearson <0.2), Moderate (0.2–0.5), and High (>0.5)—without **(e)** and with **(f)** Supervised-PCA. The white point in all the boxplots marks the mean Spearman correlation value.

For BLL architectures, One-Hot encoding was most frequently associated with top-performing models, accounting for 7 of the 27 datasets (Supplementary Figure 2a). ProtT5 (6), ESM2-650 (3), and ANKH2-E1 (2) were the next most frequent top-performing representations. Overall, One-Hot encoding showed predictive performance comparable to larger pLM embeddings (Figure **4**a). Following SPCA dimensionality reduction, the performance of most representations improved (Figure **4**b). Under these conditions, ANKH3-XL, and ESM2-3B emerged among the strongest performers, whereas One-Hot encoding showed little benefit from dimensionality reduction. Despite this, One-Hot remained one of the most competitive representations, with five top-performing models, second only to ANKH2-E1 (7) (Supplementary Figure 2b). Overall, dimensionality reduction increased the predictive performance of most representations across all architectures (Figure **4**).

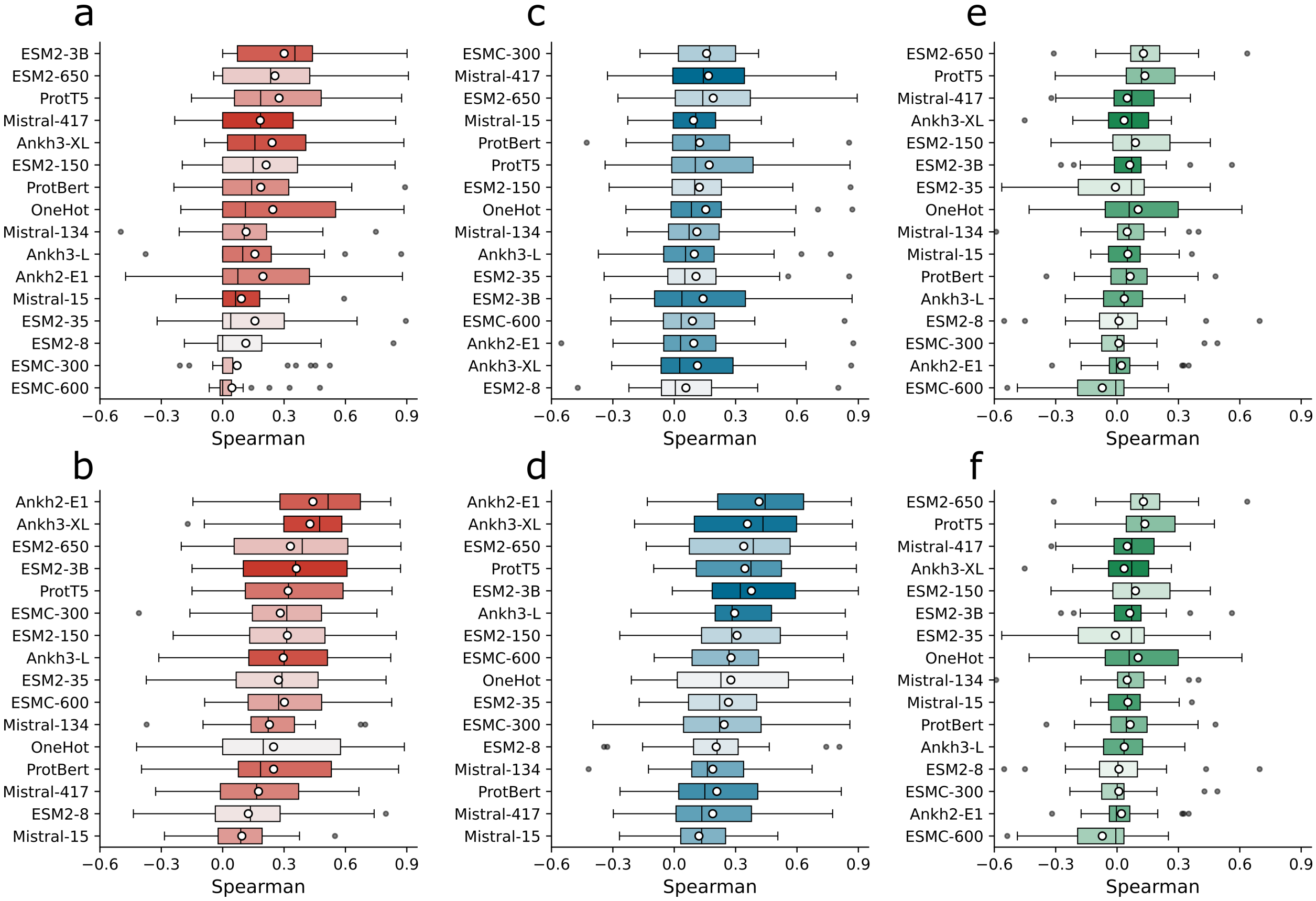

**Figure 4**: **Performance analysis of numerical representations across Multitask Bayesian Neural Network architectures.** Predictive performance of different numerical representations evaluated across benchmarking datasets using Bayesian Last Layer (BLL, red), Monte Carlo Dropout (MCD, blue), and Hierarchical Bayesian Networks (HBN, green). **a, c, e** show the distribution of Spearman correlation coefficients obtained without dimensionality reduction for BLL, MCD, and HBN architectures, respectively. **b, d, f** show the corresponding performance distributions after Supervised-PCA dimensionality reduction. Boxplots summarize the distribution of Spearman correlation coefficients across all datasets and numerical representations, with white dots indicating the mean value in each distribution.

## 2.5 Calibration performance of the Bayesian neural network architectures

To evaluate the reliability of the probabilistic predictions produced by each Bayesian architecture, we compared the expected calibration error (ECE) across regimes. ECE measures the agreement between the predicted confidence and the observed empirical accuracy [25]. An ECE of 0 indicates perfect calibration, whereas larger values indicate increasing miscalibration.

Across all dataset-size and label-correlation regimes, BLL consistently exhibited the best calibration, with ECE values of 0.4 or lower (Figure **5**a,b). Moreover, calibration improved as dataset size increased, with ECE decreasing from the Low to Moderate and from the Moderate to High data regimes (Figure **5**a). In contrast, HBN displayed the poorest calibration, with ECE values ranging from 0.6 to 0.8, significantly higher than those of BLL across all regimes. No significant differences in calibration were observed between BLL and MCD, although both architectures achieved their lowest ECE values in the High data-size and High label-correlation regimes.

Applying SPCA dimensionality reduction substantially improved the calibration of HBN across both dataset-size and label-correlation regimes (Figure **5**c,d; $P$ <0.01). In the Moderate and High data regimes, HBN achieved ECE values comparable to those of BLL and MCD, although these differences were not statistically significant. In contrast, SPCA had little effect on the calibration of BLL and MCD, as no significant changes in ECE were detected across any dataset-size or label-correlation regime before and after dimensionality reduction.

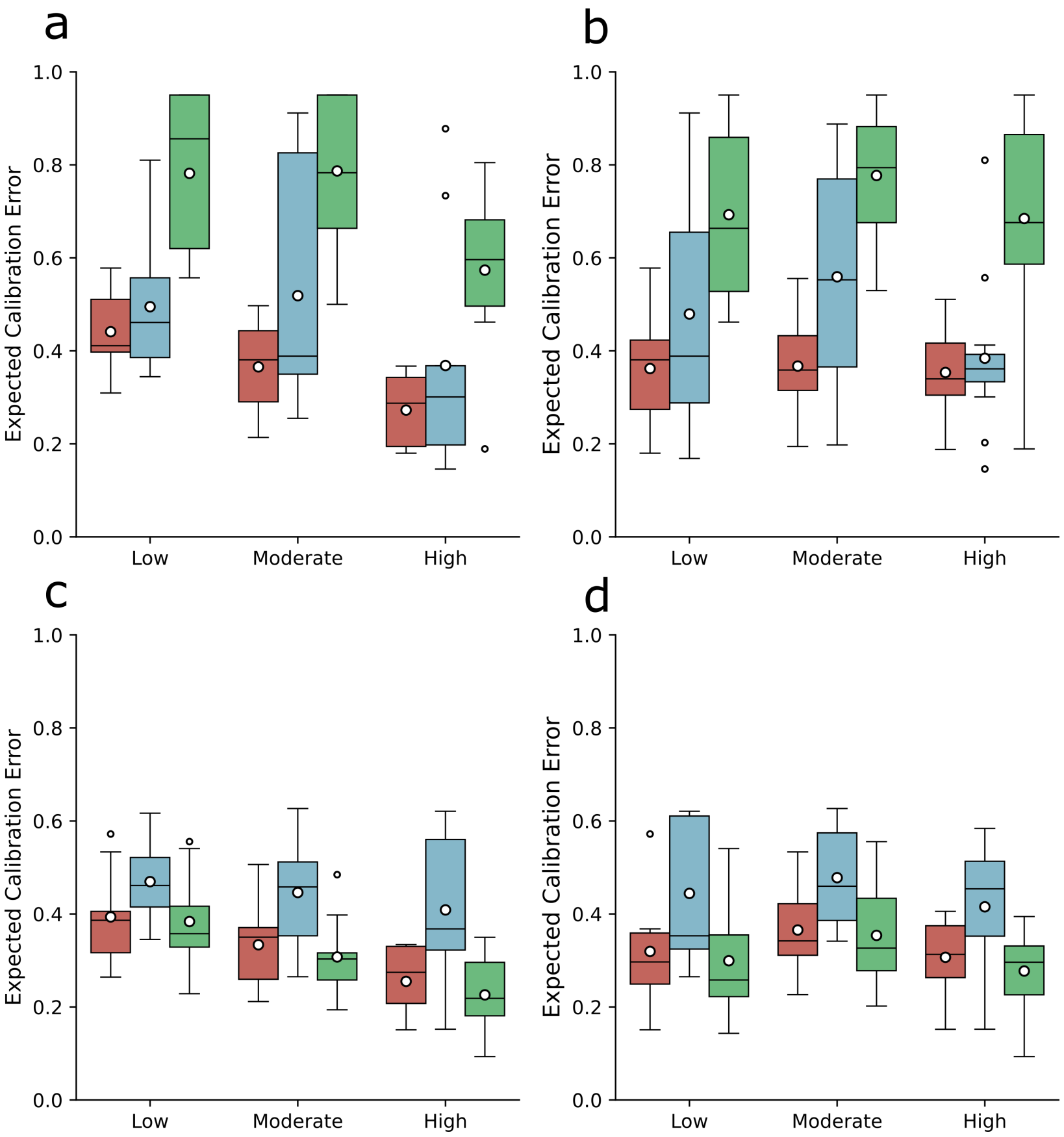

**Figure 5**: **Calibration performance of Bayesian neural network architectures across dataset-size and label-correlation regimes.** Distribution of the expected calibration error (ECE) for Bayesian Last Layer (BLL, red), Monte Carlo Dropout (MCD, blue), and Hierarchical Bayesian Networks (HBN, green) across the benchmark datasets. **a, c** ECE stratified by dataset size—Low (<100 entries), Moderate (100–1000 entries), and High (>1000 entries)—without **(a)** and with **(c)** Supervised-PCA dimensionality reduction. **b, d** ECE stratified by inter-property label correlation—Low (Pearson <0.2), Moderate (0.2–0.5), and High (>0.5)—without **(b)** and with **(d)** Supervised-PCA dimensionality reduction. Boxplots summarize the distribution of Spearman correlation coefficients across all datasets and numerical representations, with white dots indicating the mean value in each distribution.

### 2.6 Dataset specific performance: Rhamnosyl Transferase 1 Chain A

To investigate how the different Bayesian architectures perform on individual prediction tasks, we selected the Rhamnosyl Transferase 1 Chain A (RhlA) benchmark [26]. Without dimensionality reduction, activity was predicted slightly more accurately than selectivity across all architectures (Figure **6**a). Among the three Bayesian approaches, BLL consistently achieved the highest Spearman correlation coefficients for both properties, whereas HBN showed the lowest predictive performance. Following SPCA dimensionality reduction, the performance differences between architectures became substantially smaller, resulting in comparable predictive accuracy across both activity and selectivity (Figure **6**b).

We next examined the influence of the sequence numerical representation on multitask performance. For the combined prediction of activity and selectivity, OneHot and ESM2-150 achieved the highest average performance before dimensionality reduction (Figure **6**c). Applying SPCA led to a marked improvement across all representations, with ESM2-35 and ESM2-8 becoming the top-performing representations (Figure **6**d).

A similar analysis performed separately for each property revealed distinct trends before dimensionality reduction. For activity prediction, OneHot and Mistral-417 produced the highest Spearman correlations (Figure **6**e), whereas ESM2-35 and ESM2-150 performed best after SPCA (Figure **6**f). For selectivity prediction, OneHot and ESM2-150 were the highest-performing representations before dimensionality reduction (Figure **6**g), while ESM2-8 and ProtBert achieved the best performance after SPCA (Figure **6**h), with more stability after SPCA dimensionality reduction. This increased stability was observed for the joint prediction of activity and selectivity as well as for each property individually, and the same trend was reproduced across the additional benchmark datasets (Supplementary Section 4).

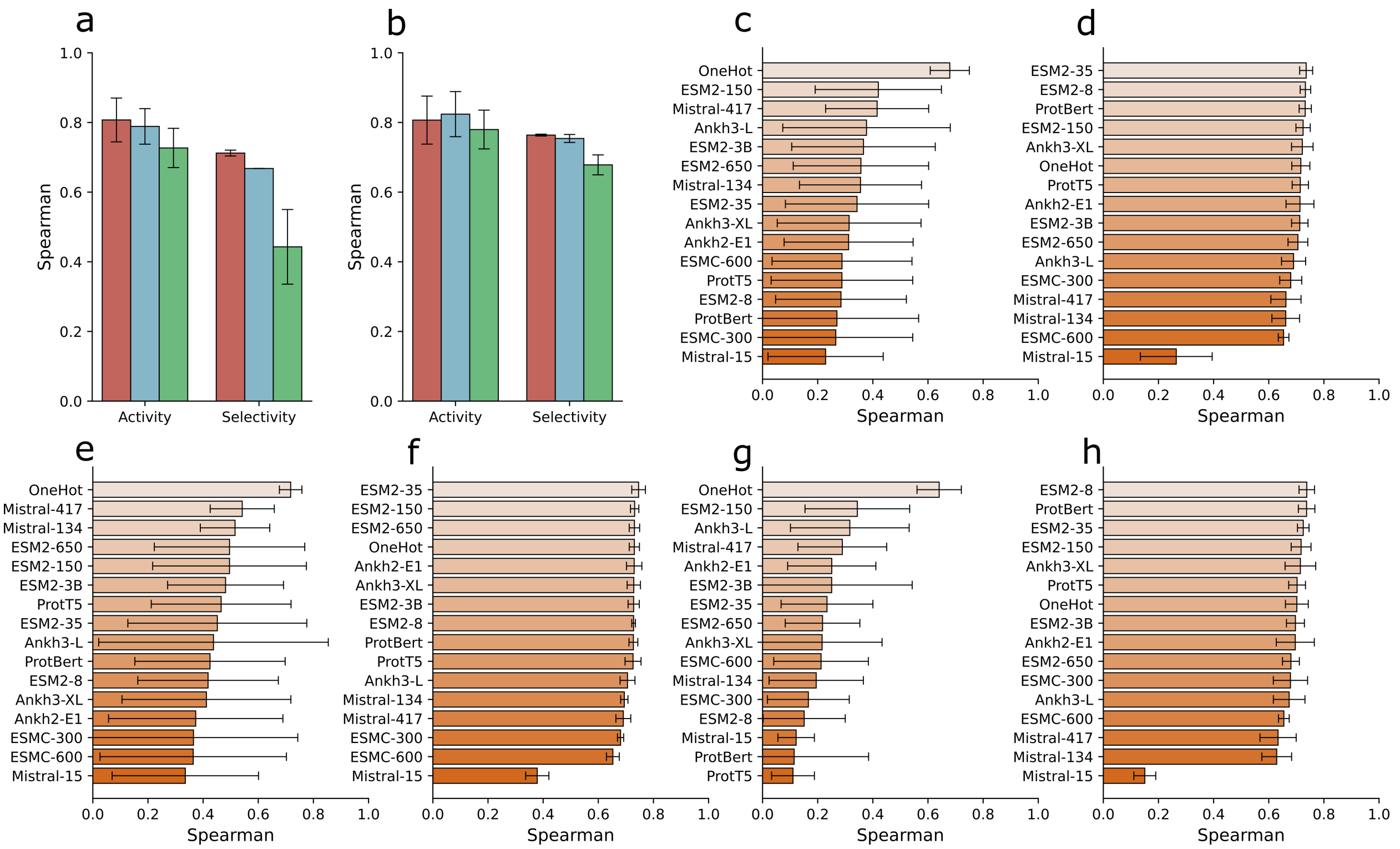

**Figure 6**: **Predictive performance on the Rhamnosyl Transferase 1 Chain A benchmark. a, b** Spearman correlation coefficients obtained by the three best-performing numerical representations for Bayesian Last Layer (BLL, red), Monte Carlo Dropout (MCD, blue), and Hierarchical Bayesian Neural Networks (HBN, green) for the joint prediction of activity and selectivity, evaluated without **(a)** and with **(b)** Supervised-PCA dimensionality reduction. **c, d** Average Spearman correlation coefficients across the three Bayesian architectures for the combined activity and selectivity prediction, evaluated without **(c)** and with **(d)** Supervised-PCA dimensionality reduction. **e, f** Spearman correlation coefficients for activity prediction, averaged across the three Bayesian architectures, evaluated without **(e)** and with **(f)** Supervised-PCA dimensionality reduction. **g, h** Spearman correlation coefficients for selectivity prediction, averaged across the three Bayesian architectures, evaluated without **(g)** and with **(h)** Supervised-PCA dimensionality reduction.

## 3 Discussion

Our large-scale benchmark of BNNs revealed several consistent and previously underexplored trends in multiparameter protein engineering. First, BLL architecture (partial Bayesian parametrization), consistently achieved the strongest balance between predictive accuracy, generalization, and uncertainty estimation across diverse datasets and sequence representations. This finding is notable because recent studies have explored more extensively parametrized Bayesian architectures due to their potential to improve robustness and uncertainty estimation in Low-data settings [27–32]. However, accurately estimating uncertainty in highly parameterized Bayesian models becomes increasingly challenging when the amount of training data is small relative to the dimensionality of the feature space [29]. Consistent with this interpretation, the comparatively weaker performance of HBN, indicates that increasing Bayesian parameterization alone does not necessarily improve predictive performance.

The consistently low overfitting observed for BLL models further supports this interpretation. These models showed only small differences between validation and test performance, even in limited data regimes, indicating strong generalization. More broadly, all evaluated architectures exhibited relatively low overfitting before dimensionality reduction, a behavior previously reported for BNNs [27]. Our results extend these observations to multiparameter protein engineering tasks. Bayesian priors regularize model parameters and explicitly account for uncertainty during training [10, 11, 33]. Together with optimization strategies such as dropout and early stopping [34, 35], these mechanisms likely limited excessive memorization despite the high dimensionality of pLM embeddings.

Dataset size alone did not fully explain predictive performance or generalization capacity. Notably, several small datasets achieved predictive performance comparable to that of substantially larger datasets. These differences likely reflect additional factors, such as data quality, signal-to-noise ratio, assay variability, and the complexity of the protein fitness landscape [36, 37]. These findings emphasize that dataset quality may be more important determinants of predictive performance than dataset size alone, as previously reported [37–39].

A second major finding was the consistent benefit of SPCA. Across most architectures, SPCA substantially improved predictive performance, with MCD models showing the largest relative gains. SPCA also markedly improved uncertainty calibration, particularly for HBN, whose expected calibration error became comparable to that of BLL and MCD after dimensionality reduction. This result suggests that pLM embeddings contain substantial redundancy and correlated latent features [40]. Compressing these representations likely improves both optimization and uncertainty estimation by removing redundant components that contribute little to task-relevant prediction [40].

Although SPCA generally improved performance, its effects differed across architectures. BLL models remained among the strongest overall performers but exhibited increased overfitting following dimensionality reduction. One possible explanation is that Bayesian regularization in BLL is confined to the final layer [13, 14], making the preceding deterministic layers more sensitive to information loss introduced during feature compression. Because PCA removes low-variance components independently of their predictive relevance, some discarded features may still contain complementary biological information [40]. In contrast, MCD and HBN introduce stochastic regularization throughout multiple network layers [12, 15].

Additional evidence supporting the presence of substantial redundancy in pLM embeddings came from the remarkable consistency in the optimal number of principal components. Across datasets and sequence representations, the best-performing models generally required a number of principal components corresponding to only approximately 10–12% of the available training samples (Supplementary Fig. 3). Furthermore, predictive performance showed little relationship with the proportion of explained variance retained after compression. This observation suggests that preserving global variance is not equivalent to preserving task-relevant biological information. Similar conclusions have recently been reported for pLMs[40]. Consistent with this interpretation, the largest improvements following SPCA were observed for the highest-dimensional pLM embeddings, including ANKH3-XL, ANKH2-E1, ESM2-3B, and ESM2-650. Because these representations originate from models containing between 650 million and 5.7 billion parameters [41, 42], they likely encode richer but also substantially more redundant latent information[40].

Another unexpected finding was the consistently strong performance of One-Hot encoding. Despite lacking evolutionary pretraining and structural information captured by pLMs, One-Hot representations repeatedly achieved predictive performance comparable to state-of-the-art protein language models and even outperformed several pLM embeddings in high-data regimes. This finding contrasts with the prevailing assumption that pLMs generally outperform simpler sequence representations in protein fitness prediction [43–46]. Our results suggest that, when sufficient experimental data are available, simple sequence encodings can remain highly competitive. The mechanisms underlying this observation remain unclear and may involve differences in representation sparsity, inductive bias, latent geometry, or interactions with Bayesian learning, all of which require further investigation.

We also found that BNNs maintained strong predictive performance across datasets spanning a wide range of label-correlation regimes. This observation contrasts with the common expectation that multitask learning primarily benefits highly correlated outputs [6, 8, 47]. Instead, our results indicate that BNNs can effectively exploit partially independent phenotypic signals and are not restricted to strongly coupled prediction tasks. Notably, some of the highest predictive performances were obtained for datasets with relatively weak label correlations, particularly for BLL and MCD architectures. Nevertheless, because our benchmark did not explicitly separate biological relationships from dataset-specific experimental effects, the mechanisms underlying these observations remain to be clarified. Additionally, direct benchmarking with single task models is still required.

The benchmark datasets containing multiple experimentally measured properties provided additional support to the role of dimensionality reduction. Across activity and selectivity prediction in the Rhamnosyl Transferase 1 Chain A dataset [26], different sequence representations were optimal for different properties before dimensionality reduction. After SPCA, however, performance differences between representations and Bayesian architectures became substantially smaller, resulting in more stable predictive performance across tasks. The same trend was consistently reproduced in the additional benchmark datasets. Such robustness is particularly advantageous in multiparameter protein engineering, where models are expected to generalize across multiple objectives simultaneously.

Together, these findings demonstrate the strong potential of BNNs for multiparameter protein engineering while providing practical guidelines for their application. Rather than favoring increasingly complex Bayesian formulations or ever-larger pLMs, our results show that predictive performance emerges from the interplay between Bayesian parameterization, sequence representation, feature dimensionality, and dataset characteristics. Across diverse learning regimes, BLL architectures consistently provided the most favorable balance between predictive accuracy, generalization, and calibrated uncertainty, generally enhanced when using SPCA dimensionality reduction. To the best of our knowledge, this work represents the first systematic benchmark of BNNs for multiparameter protein engineering and identifies supervised dimensionality reduction as a simple yet highly effective strategy for improving learning efficiency in high-dimensional protein representations. These findings provide a foundation for the development of more robust, data-efficient, and generalizable ML frameworks for protein engineering and multiparameter optimization.

# 4 Methods

## 4.1 Data collection and preprocessing

Multiparameter protein variant datasets containing 2 or more experimentally measured protein properties were collected from research articles and databases (See Supplementary Table 1). During preprocessing, noninformative and redundant columns were removed, and entries missing sequence information or one or more target labels were excluded from further analysis. Experimental labels were normalized relative to the corresponding wild-type reference measurement using assay-specific transformations, such as relative differences and logarithmic scaling to reduce skewness and stabilize variance.

Sequence preprocessing included reconstruction of full-length mutant sequences from mutation annotations using the corresponding wild-type reference sequence, generation of standardized mutant sequences, and creation of unique mutant identifiers encoding amino acid substitutions relative to the wild-type sequence. All processing operations were programatically logged, and summary reports were generated describing removed entries and columns, label transformation, and sequence-processing steps (see Supplementary File 1).

Finally, a machine-readable dataset catalog in JSON format was generated to facilitate FAIR data management principles. This catalog contains the metadata of each dataset, including UniProt or PDB identifiers, link to the original data source, associated publication, and description of the experimental labels (see Supplementary File 2).

## 4.2 PCA-based data split

For model training optimization and evaluation, datasets were partitioned using a deterministic PCA-guided splitting strategy designed to reduce data leakage between subsets. Protein variant sequences were first represented using 41 global physicochemical descriptors derived from the amino acid sequence properties. The resulting feature matrix was standardized and projected into a PCA space. Samples were ordered according to their coordinates along the first principal component (PC1), which captures the largest source of explained variance in the feature distribution.

Train, validation, and test subsets were subsequently assigned along the order PC1 axis using a 70:10:20 ratio. This strategy preserves the underlying structure of the sequence feature space while reducing the likelihood that highly similar variants are randomly distributed across subsets, thereby providing a more objective assessment of model generalization performance. The first 2 PCs and their explained variance ratios were retained for visualization and characterization of the feature-space structure.

### 4.3 Protein sequence representations

Protein sequences were represented using a diverse collection of pretrained pLMs spanning a wide range of model sizes and training strategies, with parameter counts ranging from 8 million to 5.7 billion parameters. The evaluated pLMs included ESM-2 variants (8M to 3B parameters) [41], ESM-C models (300M and 600M parameters) [48], Mistral-Prot (15 to 417M parameters) [49], ProtT5 (3B parameters) [50], ProtBERT (420M parameters) [51], and ANKH models, including ANKH2-ext1, ANKH-large and ANKH-xl, (approximately 2B, 1.9B, and 5.7B parameters, respectively) [42].

For all pLMs, residue-level embeddings were aggregated using mean pooling across sequence positions, and embeddings were generated using FP32 inference precision. As baseline representation, all protein sequences were additionally encoded using one-hot amino acid representations.

### 4.4 Model architecture

We evaluated 3 MTBNNs architectures for regression (Figure **1**). The first architecture consisted of a deterministic multitask neural network with shared feature extraction backbone and task-specific output heads (Figure **1**a). To estimate predictive uncertainty, MC dropout was applied during inference time as proposed by Gal et., al [12]. Specifically, 50 stochastic forward passes were performed with dropout enabled to generate an empirical predictive distribution. The empirical mean and standard deviation across these stochastic were used as estimates of the predictive mean and uncertainty, respectively.

The second architecture was a BLL multitask neural network with a shared deterministic backbone and a probabilistic output layer (Figure **1**b), following the approach proposed by Wang et., al [13]. In this model, task predictions are generated through a Bayesian linear output layer that maps the shared representation to all task outputs simultaneously while maintaining a fully factorized Gaussian variational posterior over weights and biases.

The third architecture was a multitask HBN (Figure **1**c) that explicitly decomposes predictive uncertainty into global, task-specific, and observation-level components, following the framework proposed by Hwang et., al [15]. Predictions were obtained by summing the outputs of two Bayesian linear layers: a global Bayesian layer that models effects shared across all tasks and a task-specific Bayesian layer that captures deviations unique to each output. Both layers maintain fully factorized Gaussian variational posteriors over weights and biases with independent Gaussian priors defining the regularization strength of global and task-specific Bayesian layers and their corresponding Gaussian priors.

For both the BLL and the HBN models, predictive mean were computed as the empirical mean across 50 Monte Carlo samples, whereas epistemic uncertainty was estimated using the empirical standard deviation of the sampled. In addition, task-specific aleatoric uncertainty was modeled through learnable log-variance parameters. Epistemic and aleatoric uncertainties were summed to obtain the total predictive variance. Training of the BLL model was regularized by the Kullback–Leibler divergence between the variational posterior of the Bayesian last layer and a zero-mean Gaussian prior with fixed variance. Similarly, HBN training was regularized by the sum of Kullback–Leibler divergences between the variational posteriors of the global and task-specific Bayesian layers and their corresponding Gaussian priors.

### 4.5 Model training and hyperparameter optimization

All models were trained within a unified multitask learning pipeline implemented in PyTorch, with systematic hyperparameter optimization performed using Optuna. Early stopping and model selection were conducted exclusively on the validation set, whereas the test set was reserved for final evaluation. Training was performed for a maximum of 10000 epochs. For each Optuna trial, the hyperparameter search space was adjusted to dataset size to account for different data regimes. Specifically, progressively larger hidden dimensions and mini-batch sizes were explored for larger datasets, whereas dropout rates were reduced to balance model regularization and capacity. Learning rates were sampled on a logarithmic scale, and the degrees of freedom of the Student-t likelihood were treated as a tunable hyperparameter.

The training objective combined Student-t negative log-likelihood term with an optional Bayesian regularization term. For BLL and HBN, the loss function was defined as the mean Student-t negative log-likelihood, the Kullback–Leibler divergence between variational posteriors and their corresponding priors multiplied by a scaling factor $\beta$. For mini-batch training, $\beta$ was set proportional to the batch-to-dataset size ratio, ensuring an appropriate normalization of the Kullback–Leibler term across dataset sizes. For MCD, the Kullback–Leibler term was omitted.

Final models were retrained from scratch using the optimal hyperparameters identified during the Optuna optimization and subsequently evaluated on the held-out test set. To evaluate the effect of dimensionality reduction, all sequence representations were additionally transformed using SPCA [17], with cross-validated selection of the optimal number of retained principal components. The evaluated dimensionalities corresponded to fractions of the training set size, specifically $n/6$, $n/4$, $n/3$, $n/2$, $0.66n$, $0.75n$ and $n-1$, where $n$ denotes the number of training samples. This strategies enabled systematic evaluation of different compression levels while ensuring that the number of retained principal components remained below the number of training samples, thereby improving statistical stability and reducing the risk of overparametrization in Low-data regime.

### 4.6 Model evaluation

Model performance was primarily evaluated using the Spearman correlation coefficient, computed independently for each prediction task across tasks to obtain global performance estimate for each dataset. Spearman correlation was selected because it measure monotonic rank-based associations between predicted and experimentally observed values, thereby assessing the ability of the models to correctly rank protein variants independently of absolute prediction scale [18]. Similarly, ECE[25] was used to measure the calibration of the models.

All evaluations were performed separately on the validation and test sets using model predictions obtained after model convergence. Generalization performance was assessed by quantifying the validation-test performance across architectures, dataset-size regimes, and label-correlation categories, with larger discrepancies indicating increased overfitting and reduced generalization capacity.

### 4.7 Statistical Analysis

Statistical analyses were performed using nonparametric tests to evaluate differences in predictive performance across sequence representations, model architectures, dimensionality-reduction strategies, dataset-size categories, and label-correlation regimes. Global comparisons involving paired multi-group analyses were conducted using Friedman tests, whereas Kruskal–Wallis tests were used for comparisons between independent groups. Pairwise comparisons were performed using Wilcoxon signed-rank or Mann–Whitney U tests, depending on whether samples were paired or independent.

Multiple hypothesis testing was corrected using the Benjamini–Hochberg false discovery rate (FDR) procedure. Effect sizes were quantified using rank-biserial correlation and Cohen's $d$, where appropriate. Performance differences were further summarized using percentage performance improvements, 95% confidence intervals, and descriptive statistics, including means, medians, and standard deviations. Statistical significance was defined as a two-sided adjusted p-value ($P < 0.05$). Significance levels were defined as follows: * $P < 0.05$, ** $P < 0.01$, and *** $P < 0.001$; ns, not significant. False discovery rate (FDR) correction was applied for multiple comparisons where indicated. The results of the statistical analysis are provided in Supplementary File 3 and 4 for Spearman correlation coefficient and ECE, respectively.

## Data availability

The datasets generated and analyzed during this study will be publicly available through GitHub (`https://github.com/ipb-halle/MTLMPO_WyrzykalaProject`) and Zenodo (`https://zenodo.org/records/21512415`) upon publication. The released resources include 27 multiparameter protein engineering datasets together with their corresponding catalogs containing metadata and traceability information describing the origin, processing, and organization of the data.

## Code availability

Code used for this study is available via GitHub at `https://github.com/ipb-halle/MTLMPO_WyrzykalaProject`.

## Conflict of interest statement

The authors declare no conflict of interests.

## Author contributions statement

F.H-R. curated and cleaned the data; designed, implemented, and trained the neural networks; performed the data analysis; generated the figures; and wrote the original manuscript draft. D.M-O. contributed to data analysis, code review, and manuscript writing. D.W. and T.S.S. contributed to data collection and curation, as well as model training and data analysis. F.H-R., D.M-O., and M.D.D. conceived and conceptualized the study. M.D.D. supervised the research and acquired funding. All authors reviewed, edited, and approved the final manuscript.

## Acknowledgments

We acknowledge primarily funding for this research by the BMFTR (Förderkennzeichen 031B1442E and 031B1449C). M.D.D. acknowledges funding by the Deutsche Forschungsgemeinschaft (DFG, German Research Foundation) – within the Priority Program Molecular Machine Learning SPP2363 (Project Number 497207454), BMFTR (Förderkennzeichen 031B1442A). We also thank Dr. Steffen Neumann and the German Network for Bioinformatics Infrastructure (de.NBI) for the computational resources provided to this project. Authors thank Prof. Michael A. Nash and Peter Zaspel for their critical reading of the manuscript.

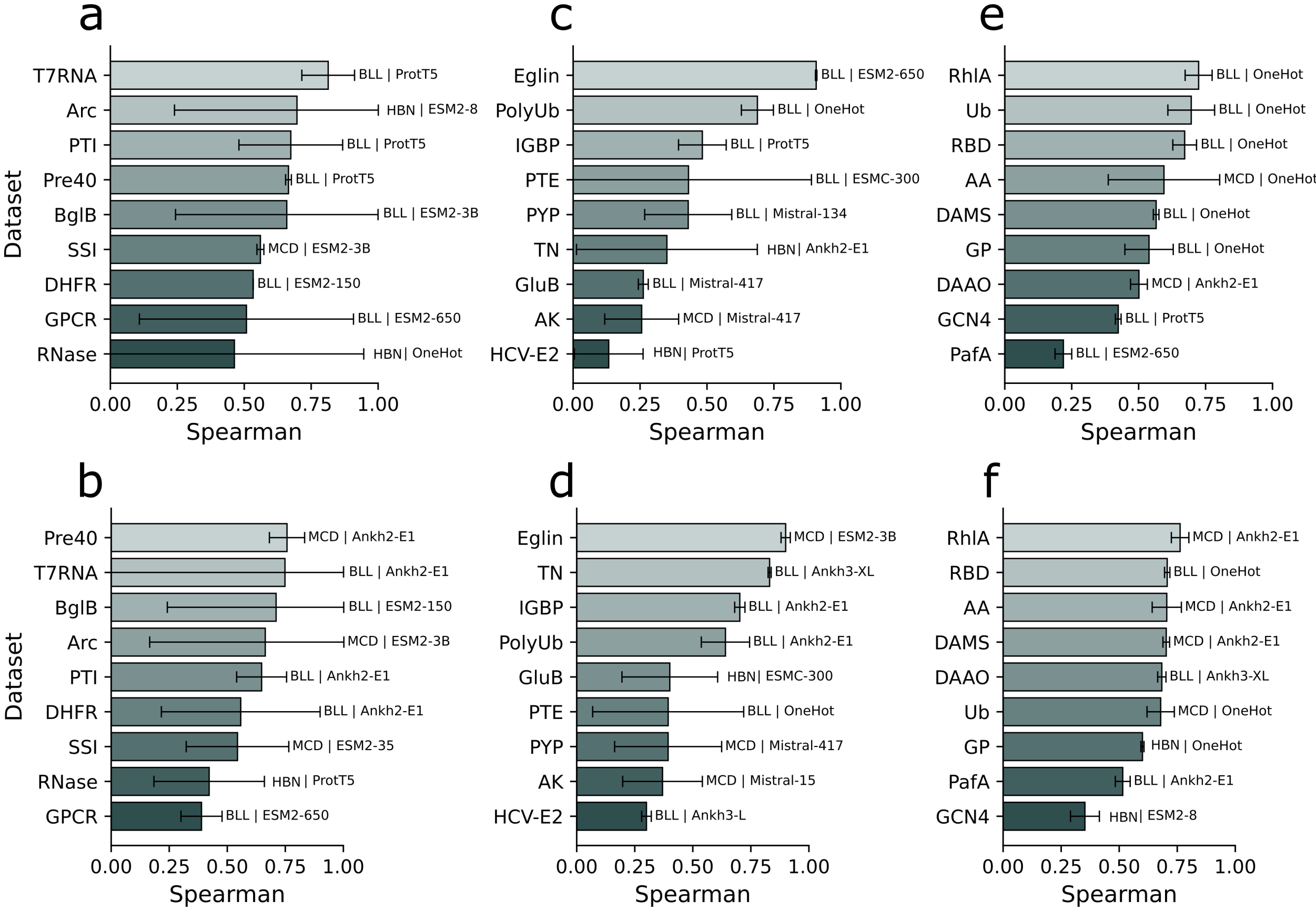

**Extended Data Figure 1**: **Best-performing model and numerical representation across benchmark datasets stratified by dataset regime.** Horizontal bar plots summarize the highest test Spearman correlation achieved for each protein mutational dataset after selecting the best-performing combination of machine learning model and numerical representation. Bar lengths represent the test Spearman correlation of the selected model, and error bars indicate the standard deviation between the validation and test Spearman correlations for the corresponding model–representation combination, providing an estimate of overfitting. Text annotations to the right of each bar identify the selected machine learning model and numerical representation. Dataset names are abbreviated for clarity. The full name of the datasets are in Supplementary Table 1. **a, b** Performance of Low-size datasets (<100 entries), without **(a)** and with **(b)** Supervised-PCA dimensionality reduction. (**c, d**) Performance of Moderate-size datasets (100–1000 entries), without **(c)** and with **(d)** Supervised-PCA. (**e, f**) Performance of High-size datasets(>100 entries)—without **(e)** and with **(f)** Supervised-PCA. Bayesian Last Layer (BLL), Monte Carlo Dropout (MCD), and Hierarchical Bayesian Neural Networks (HBN) across datasets.

# Supplementary Information

# Multitask Bayesian Neural Networks for Multiparameter Protein Engineering

**Fabio Herrera-Rocha[1], David Medina-Ortiz[1], Desiree Wyrzykala[1], Tharun Srinivasan Sudha[1], and Mehdi D. Davari[1*]**

[1]Leibniz-Institute of Plant Biochemistry, Department of Bioorganic Chemistry, Weinberg 3, 06120 Halle, Germany

## Contents

## List of Figures

*mehdi.davari@ipb-halle.de

## List of Tables

## S1 Datasets Overview

The collection of benchmark datasets spans 27 protein systems and includes both single-point and mixed mutagenesis experiments (Supplementary Table 1). Dataset sizes vary by several orders of magnitude, ranging from small-scale assays with 41 variants to large-scale landscapes containing 12294 variants. This wide range enables systematic evaluation of model performance in both low-data regimes—where Bayesian methods are expected to provide strong benefits through uncertainty quantification[1]—and high-data regimes where expressive function approximation becomes critical.

The datasets also differ in the number of measured functional properties, ranging from 2 objective assays to highly multi-dimensional phenotypic readouts (up to 16 properties for the Hepatitis C virus E2 glycoprotein). Related phenotypes may share underlying biophysical constraints, enabling information sharing across tasks and improved sample efficiency.

Finally, the datasets exhibit substantial variation in the correlation between functional properties, as reflected by Pearson correlation coefficients ranging from near-zero or weak signal regimes (e.g., $\beta$-glucosidase B) to strong predictive structure (e.g., D-amino acid oxidase multisubstrate, Ubiquitin, and Transcriptional repressor arc). This heterogeneity provides a rigorous benchmark for evaluating not only predictive accuracy but also the calibration and robustness of uncertainty estimates produced by Bayesian neural networks.

**Supplementary Table 1**: **Summary of the protein mutational datasets used for benchmarking.** The table lists the 27 protein mutational datasets included in this study together with the mutation type (single-point or mixed mutations), dataset size (number of variants), number of experimentally measured properties, average pairwise Pearson correlation coefficient between measured properties, and the corresponding original publication. Dataset sizes span from 41 to 12,294 variants and include between two and sixteen measured properties, providing a diverse benchmark across different data regimes and levels of inter-property correlation.

| Dataset | Abbreviation | Mutation Type | Size | Properties | Pearson Correlation | Reference |
|---|---|---|---|---|---|---|
| Pancreatic trypsin inhibitior | PTI | Single-point | 41 | 2 | 0.753 | Yu et al.,[2] |
| T7 RNA polymerase | T7RNA | Mixed | 42 | 2 | 0.297 | Boulain et al.,[3] |
| Transcriptional repressor arc | Arc | Single-point | 46 | 4 | 0.792 | Milla et al.,[4] |
| Pre-mRNA | Pre40 | Mixed | 47 | 2 | 0.253 | Jemth et al.,[5] |
| Dihydrofolate reductase | DHFR | Mixed | 51 | 3 | 0.784 | Arai & Iwakura[6] |
| Ribonuclease | RNase | Mixed | 65 | 3 | 0.612 | Serrano et al.,[7] |
| Adhesion G protein-coupled receptor | GPCR | Single-point | 66 | 2 | 0.632 | Nazarko et al.,[8] |
| Subtilisin-chymotrypsin inhibitor 2A | SSI | Mixed | 67 | 3 | 0.501 | Itzhaki et al.,[9] |
| Glucosidase B | GluB | Single-point | 78 | 2 | -0.204 | d2dcure.com |
| Thermonuclease | TN | Single-point | 102 | 2 | 0.509 | Meeker et al.,[10] |
| Photoactive yellow protein | PYP | Single-point | 104 | 2 | 0.043 | Philip et al.,[11] |
| Adenylate kinase | AK | Mixed | 105 | 6 | 0.046 | Howell et al.,[12] |
| Bacterial phosphotriesterase | PTE | Mixed | 124 | 2 | -0.180 | Kaltenbach et al.,[13] |
| Eglin C | Eglin | Mixed | 300 | 2 | -0.068 | Yi et al.,[14] |
| Hepatitis C virus E2 glycoprotein | HCV-E2 | Mixed | 309 | 16 | 0.440 | Pierce et al.,[15] |
| Polyubiquitin-C | PolyUb | Mixed | 332 | 2 | 0.656 | Roscoe et al.,[16] |
| $\beta$-glucosidase B | BglB | Single-point | 398 | 2 | 0.021 | Carlin et al.,[17] |
| Immunoglobulin G-binding protein G | IGBP | Single-point | 927 | 2 | -0.446 | Olson et al.,[18] |
| Alkaline Phosphatase PafA | PafA | Single-point | 1039 | 3 | 0.341 | Markin et al.,[19] |
| Ubiquitin | Ub | Mixed | 1157 | 5 | 0.845 | Mavor et al.,[20] |
| Genome polyprotein | GP | Mixed | 1631 | 2 | 0.599 | Qi et al.,[21] |
| Rhamnosyl Transferase 1 Chain A | RhlA | Mixed | 2518 | 2 | -0.106 | Fu et al.,[22] |
| General control protein | GCN4 | Mixed | 2760 | 3 | -0.161 | Staller et al.,[23] |
| SARS-CoV-2 RBD | RBD | Single-point | 3998 | 2 | 0.654 | Starr et al.,[24] |
| D-amino acid oxidase multisubstrate | DAMS | Single-point | 5800 | 5 | 0.951 | Vanella et al.,[25] |
| D-Amino acid oxydase | DAAO | Single-point | 7644 | 2 | 0.543 | Vanella et al.,[26] |
| $\alpha$-Amylase | AA | Mixed | 12294 | 3 | 0.228 | van der Flier et al.,[27] |

## S2 Model performance

### S2.1 Top performing models per data and label correlation regime

To assess the consistency of each Bayesian neural network architecture across datasets, we quantified how frequently each architecture achieved the highest predictive performance (Supplementary Figure **1**). For every dataset, the best-performing configuration of each architecture was identified by selecting the run with the highest value of the evaluation metric. The architecture with the overall highest performance for that dataset was then designated as the top-performing architecture. Finally, the number of datasets for which each architecture ranked first was counted and visualized as a bar chart. This analysis provides a measure of the robustness and general applicability of each architecture by evaluating how consistently it outperformed the alternatives across diverse benchmark datasets. The results of this analysis are described in the results of the main manuscript.

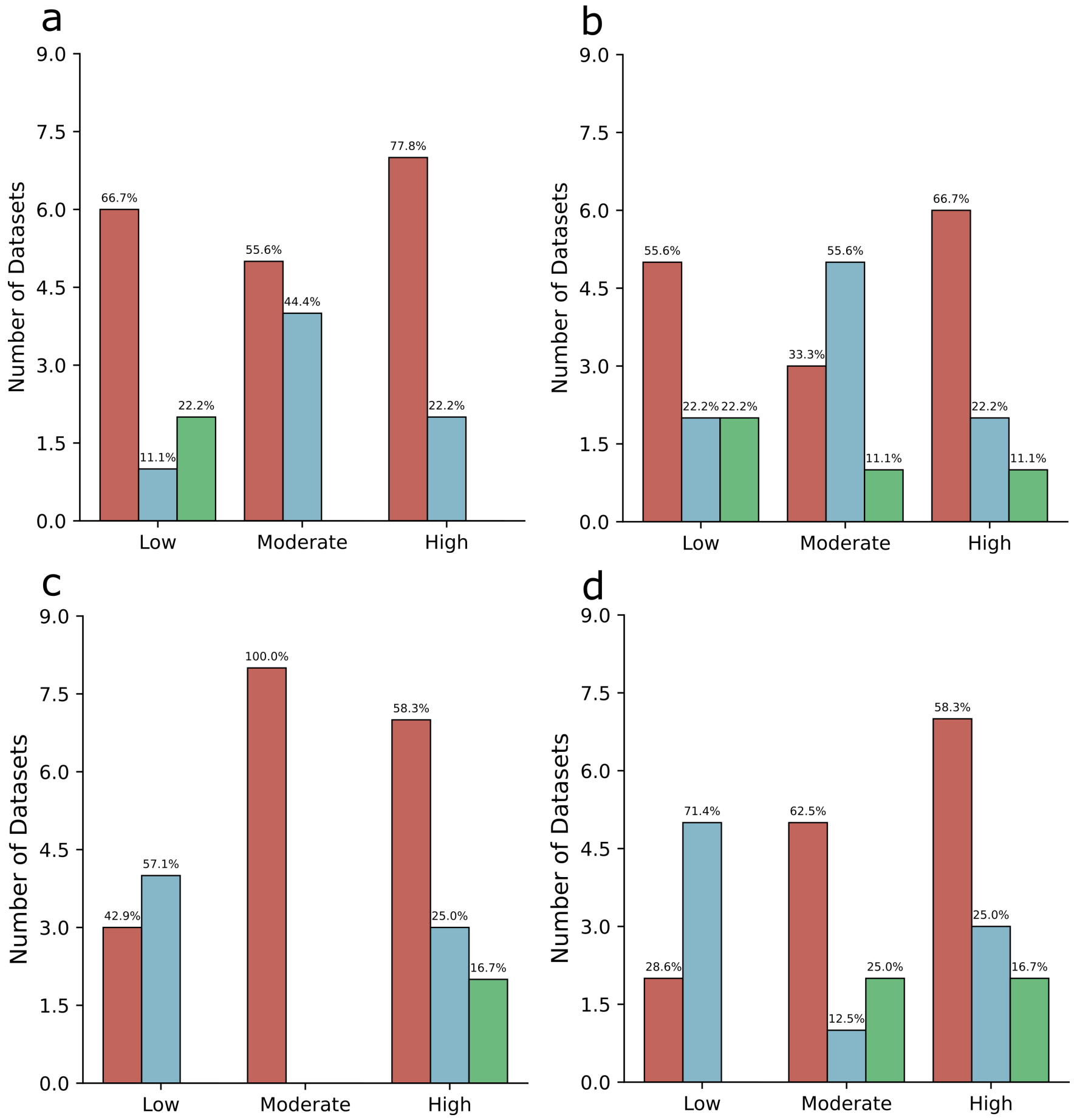

**Supplementary Figure 1**: **Top-performing model architectures across dataset size and label correlation regimes. a, b** Number of datasets for which each architecture achieved the best predictive performance across dataset size categories: Low (<100 samples), Moderate (100–1,000 samples) and High (>1,000 samples), without **(a)** and with **(b)** Supervised-PCA dimensionality reduction. (**c, d**) Number of datasets for which each architecture achieved the best predictive performance across label correlation categories: Low (Pearson correlation <0.2), Moderate (0.2–0.5) and High (>0.5), without **(c)** and with **(d)** Supervised-PC dimensionality reduction. Colours indicate the model architecture: Bayesian Last Layer (BLL, red), Monte Carlo Dropout (MCD, blue) and Hierarchical Neural Network (HBN, green).

### S2.2 Performance of the sequence representation per architecture

To assess the robustness of the numerical representations across benchmark datasets, we quantified how frequently each representation achieved the highest predictive performance. For each dataset, the mean performance of every representation was calculated across all evaluated Bayesian architectures using the selected performance metric. The representation with the highest mean performance was identified as the top-performing representation for that dataset, and the number of first-place rankings across all datasets was subsequently counted (Supplementary Figure **2**). This analysis complements the aggregate performance comparisons by evaluating how consistently a representation emerges as the best-performing option across diverse prediction tasks.

Across all Bayesian architectures, OneHot encoding consistently ranked among the three most frequently top-performing representations (Supplementary Figure 2). Prior to dimensionality reduction, ProtT5 and ESM650 also achieved a relatively high number of first-place rankings across the Bayesian Last Layer (BLL),

Monte Carlo Dropout (MCD), and Hierarchical Bayesian Neural Network (HBN) architectures (Supplementary Figure **2**a,c,e). Following dimensionality reduction, the strongest performers shifted toward the Ankh3-E1 and Ankh3-XL embeddings, which more frequently ranked first across architectures (Supplementary Figure **2**b,d,f).

Despite these trends, no single representation consistently dominated the benchmark. The most frequently top-ranked representation accounted for fewer than 30% of all datasets within every architecture, indicating that predictive performance remains dataset-dependent. These findings suggest that while certain representations exhibit greater robustness than others, no universal representation is optimal across all protein engineering tasks, further supporting the importance of evaluating multiple representation strategies when developing predictive models.

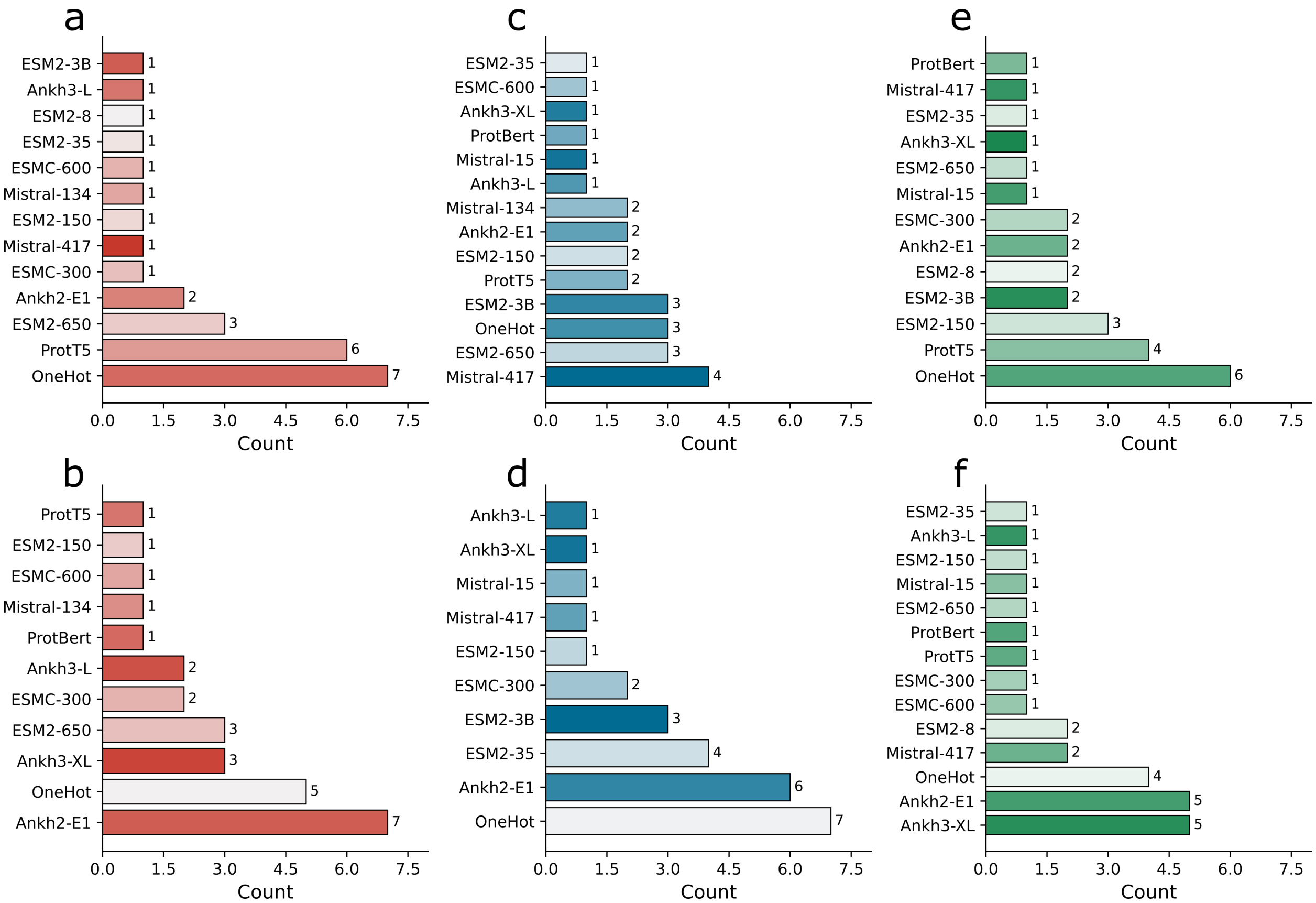

**Supplementary Figure 2**: **Count of Top-performing Numerical Representations Across Multitask Bayesian Neural Network Architectures.a, b** Count of top performing models per architecture across all numerical representations evaluated across benchmarking datasets using Bayesian Last Layer (BLL, red), Monte Carlo Dropout (MCD, blue), and Hierarchical Bayesian Networks (HBN, green). **a, c, e** show the count obtained without dimensionality reduction for BLL, MCD, and HBN architectures, respectively. **b, d, f** show the corresponding count after Supervised-PCA dimensionality reduction.

## S3 Protein sequence representation

### S3.1 General performance of the sequence representation

To further investigate the behaviour of sequence representations independently of model architecture, we compared the overall performance of the embeddings across dataset-size and label-correlation categories. Without dimensionality reduction, the best-performing representations were ESM2-3B, ESM2-650, ProtT5, One-Hot encoding, and ESM2-150 (Supplementary Figure **3**). Performance trends varied substantially across dataset sizes. In Low-data regimes, large pLM embeddings generally showed higher performance than One-Hot encoding. In contrast, in High-data regimes, One-Hot encoding achieved strong performance, surpassing several pLM-based representations (Supplementary Figure **3**a). Under Moderate-data conditions, the performance differences between representations were comparatively small. Across label-correlation categories, the top-performing representations showed largely similar predictive performance (Supplementary Figure **3**b).

After SPCA dimensionality reduction, the set of the top-performing representations changed substantially. One-Hot encoding and ESM2-150 were replaced in the top-performing group by ANKH3-XL and ANKH2-E1 (Supplementary Figure **3**c). These representations showed particularly strong performance in High-data regimes, whereas comparatively minor differences were observed between Low- and Moderate-data categories. Similarly, label-correlation analyses revealed little variation in performance across Low-, Moderate-, and High-correlation datasets (Supplementary Figure **3**d). Notably, these representations correspond to the largest pLMs evaluated in this study, with model sizes ranging from 650 million to 5.7 billion parameters, suggesting that larger pLMs may benefit most strongly from dimensionality reduction.

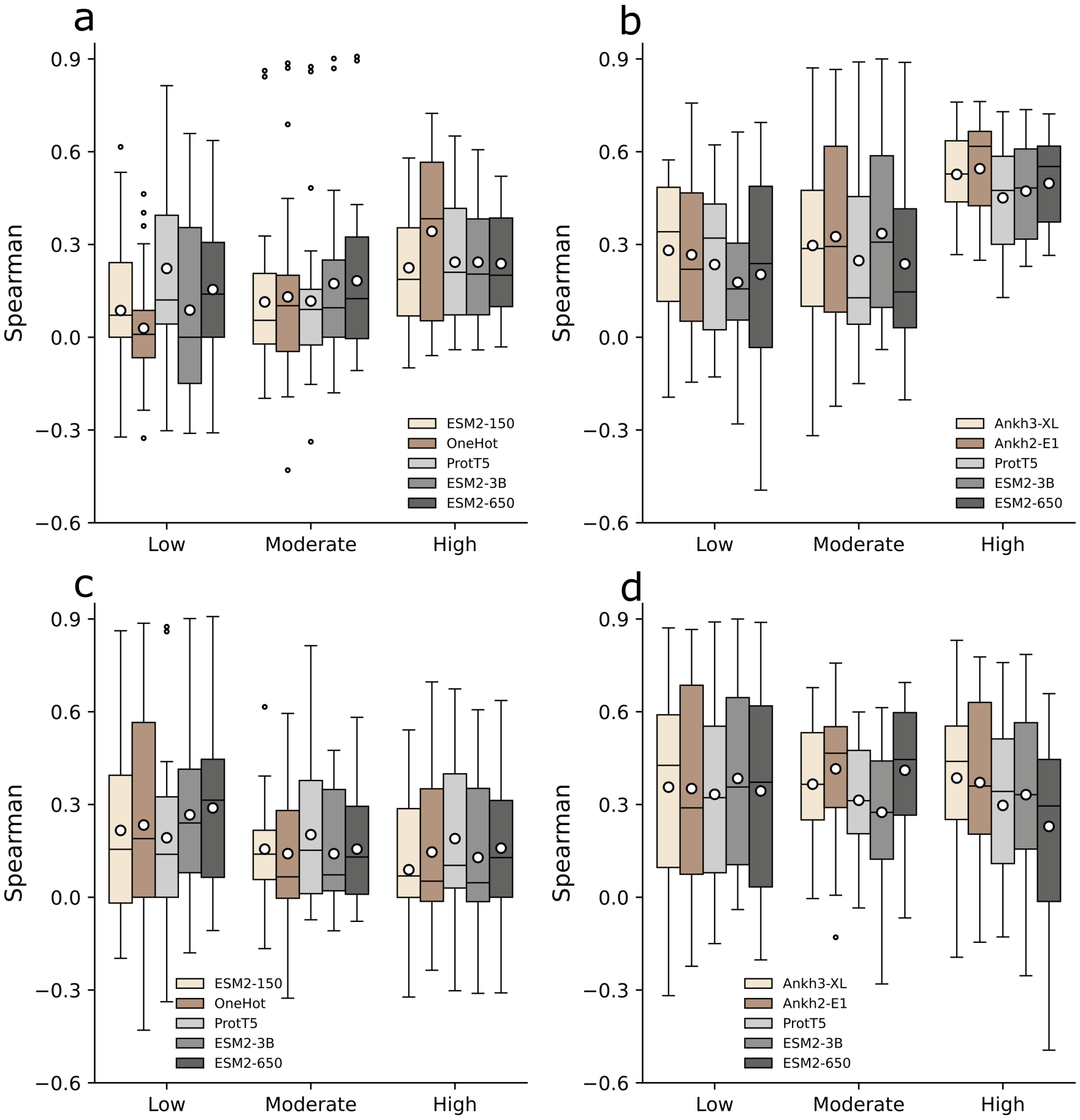

**Supplementary Figure 3**: **Performance of the Top 5 Numerical Representations across Dataset-size and Label-correlation Categories.** Predictive performance of the top-performing numerical representations evaluated across benchmarking datasets before and after Supervised-PCA dimensionality reduction. Without dimensionality reduction, the top-performing representations were ESM2-150, One-Hot, ProtT5, ESM2-3B, and ESM2-650. After dimensionality reduction, the top-performing representations were ANKH3-XL, ANKH2-E1, ProtT5, ESM2-3B, and ESM2-650. **a, b** Distribution of Spearman correlation coefficients stratified by dataset size—Low (<100 entries), Moderate (100–1000 entries), and High (>1000 entries)—without **(a)** and with **(b)** Supervised-PCA dimensionality reduction. (**c, d**) Distributions stratified by label correlation—Low (Pearson <0.2), Moderate (0.2–0.5), and High (>0.5)—without **(c)** and with **(d)** Supervised-PC dimensionality reduction. Boxplots summarize the distribution of Spearman correlation coefficients across datasets, with white dots indicating the mean value for each distribution.

### S3.2 Dimensionality reduction of the sequence representation

To investigate the behavior of feature compression across sequence representations, we analyzed the dimensionality reduction dynamics of all evaluated embeddings using SPCA[28]. In this approach, embeddings were first compressed through principal component analysis and the optimal number of principal components was subsequently selected based on the predictive performance of a multi-output ridge regressor[29]. To mitigate the curse of dimensionality [30], the number of evaluated principal components ranged from one-sixth of the training set size up to a maximum of n-1, where n corresponds to the number of training samples (see Methods for details).

Across all dataset-size and label-correlation categories, the optimal number of principal components for the models achieving the highest correspond to a relatively small fraction of the training set size, approximately 10–12% of the number of training samples (Supplementary Figure **4**a,b). This trend was highly consistent across datasets and was largely independent of the sequence representation used, indicating that substantial feature compression could be achieved without compromising predictive performance.

In contrast, the proportion of explained variance showed substantial variability across representations and datasets, ranging from approximately 40% to 99%, independently of dataset size or label-correlation regime (Supplementary Figure **4**c,d). One-Hot encoding consistently displayed the lowest explained variance values across datasets, whereas pLM embeddings generally retained more than 80% of the variance after dimensionality reduction. Despite these differences, no consistent relationship was observed between the proportion of explained variance retained and predictive performance measured by the Spearman correlation coefficient across the evaluated categories.

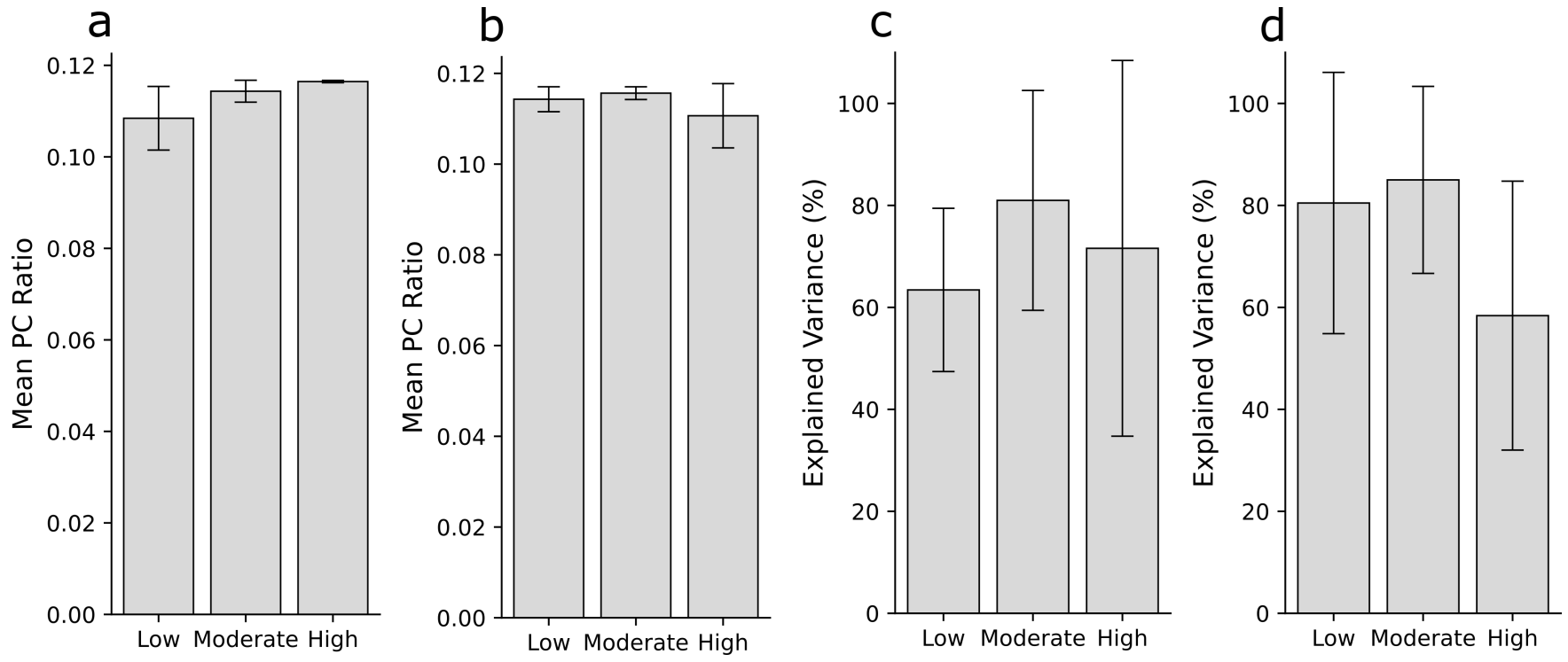

**Supplementary Figure 4**: **Analysis of Supervised-PCA Dimensionality Reduction of Embeddings for Multiparameter Protein Engineering.** Dimensionality reduction efficiency of the evaluated numerical representations after supervised-PCA dimensionality reduction for the top performing models. The feature compression was quantified as the ratio between the optimal number of principal components and the number of training examples for each embedding–dataset combination. Results are shown across dataset-size categories—Low (<100 entries), Moderate (100–1000 entries), and High (>1000 entries)—and label-correlation categories—Low (Pearson correlation <0.2), Moderate (0.2–0.5), and High (>0.5). **a, b** Mean principal component ratio stratified by dataset size **(a)** and label correlation **(b)**. (**c, d**) Mean percentage of explained variance stratified by dataset size **(c)** and label correlation **(d)**.

## S4 Performance on specific cases of study

### S4.1 General performance for the selected cases of study

To compare the overall predictive performance of the Bayesian architectures, we evaluated the best-performing numerical representations for each architecture using both the original and SPCA-dimensionality-reduced feature sets on three representative benchmark datasets: Rhamnosyl Transferase 1 Chain A (RhlA)[22], Thermonuclease[10], and the SARS-CoV-2 receptor-binding domain (SARS-CoV-2 RBD)[24] (Supplementary Figure **5**). For each architecture, the numerical representation with the highest average test Spearman correlation across the activity and selectivity prediction tasks was identified. The average Spearman correlation was used as the performance score, and the standard deviation between the two prediction tasks was included to reflect task-specific variability. This analysis summarizes the best performance by each Bayesian architecture for both properties and provides a direct comparison of their predictive capabilities before and after dimensionality reduction.

Across the three benchmark datasets, SPCA generally improved the average predictive performance of all Bayesian architectures (Supplementary Figure **5**), although the magnitude of the improvement varied between datasets. For the RhlA benchmark, predictive performance was already high with the original representations, and SPCA produced only modest additional gains across architectures(Supplementary Figure **5**a). In contrast, the Thermonuclease dataset exhibited the largest improvement following dimensionality reduction. Before SPCA, the BLL architecture achieved only limited predictive performance; however, after dimensionality reduction it became the highest-performing architecture, accompanied by substantial improvements in both MCD and HBN (Supplementary Figure **5**b). For the SARS-CoV-2 RBD dataset, BLL showed similar performance before and after SPCA, whereas MCD and HBN exhibited clear improvements after dimensionality reduction (Supplementary Figure **5**c).

Although the extent of the performance gains differed across datasets, a consistent trend emerged: dimensionality reduction either improved or maintained the predictive performance of the strongest architecture while reducing the performance gap between the Bayesian approaches. These supplementary analyses therefore reinforce the observations reported in the main text, demonstrating that the benefits of SPCA extend across multiple multitask protein engineering benchmarks while remaining dependent on the characteristics of the individual dataset.

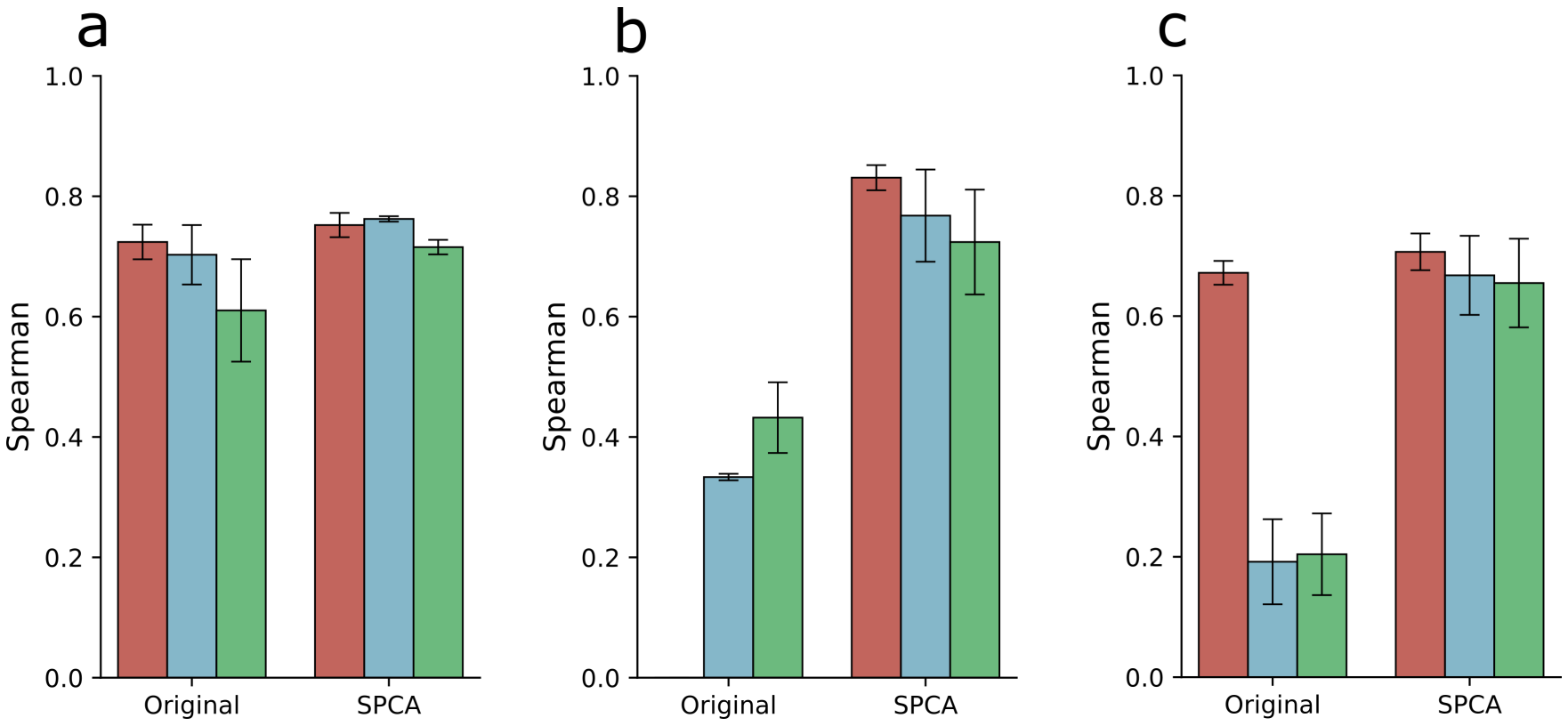

**Supplementary Figure 5**: **Predictive performance across three selected benchmark protein engineering datasets. a, b, c.** Spearman correlation coefficients obtained using the three best-performing numerical representations for Bayesian Last Layer (BLL, red), Monte Carlo Dropout (MCD, blue), and Hierarchical Bayesian Neural Networks (HBNs, green) across three benchmark datasets: Rhamnosyl Transferase 1 Chain A (**a**), Thermonuclease (**b**), and SARS-CoV-2 receptor-binding domain (**c**). Model performance is shown with and without supervised principal component analysis (SPCA) dimensionality reduction.

### S4.2 Thermonuclease benchmark

To determine whether the trends observed on RhlA generalized to other multitask protein engineering benchmarks, we performed the same analysis on the Thermonuclease[10] dataset (Supplementary Figure **6**). Unlike RhlA benchmark, the Bayesian Last Layer (BLL) architecture achieved an average predictive performance close to 0 before dimensionality reduction for the evaluated properties—Concentration of Denaturant (Cm) and Gibbs Free Energy Change (dGH2O)s— whereas MCD and HBN showed comparable but low performance (Supplementary Figure **6**a). Similar to RhlA, after SPCA, the differences between architectures were reduced, with all three approaches achieving similar Spearman correlation coefficients for the evaluated properties and BLL showing the best performance (Supplementary Figure **6**b).

Overall, the Thermonuclease benchmark reproduced the principal observations from the main text. Dimensionality reduction reduced the dependence of predictive performance on both the Bayesian architecture and the numerical representation, resulting in more consistent performance across multitask predictions. These results further support the conclusion that SPCA improves the stability of multitask learning without substantially altering the relative ranking of the Bayesian approaches.

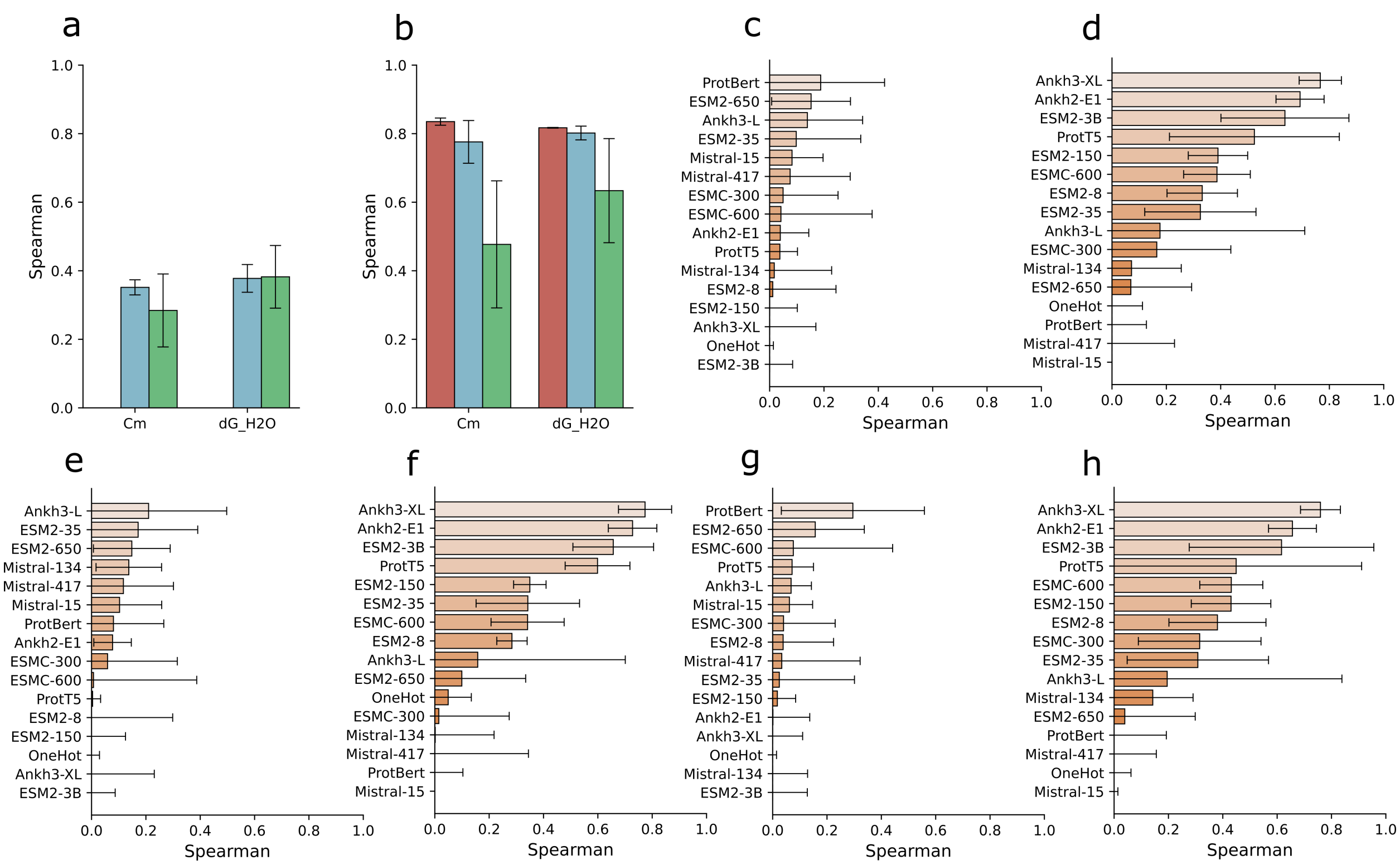

**Supplementary Figure 6**: **Predictive performance on the Thermonuclease benchmark.** **a, b** Spearman correlation coefficients obtained by the three best-performing numerical representations for Bayesian Last Layer (BLL, red), Monte Carlo Dropout (MCD, blue), and Hierarchical Bayesian Neural Networks (HBN, green) for the joint prediction of Concentration of Denaturant (Cm) and Gibbs Free Energy Change ($dG_{H2O}$), evaluated without **(a)** and with **(b)** Supervised-PCA dimensionality reduction. **c, d** Average Spearman correlation coefficients across the three Bayesian architectures for the combined Cm and $dG_{H2O}$ prediction, evaluated without **(c)** and with **(d)** Supervised-PCA dimensionality reduction. **e, f** Spearman correlation coefficients for Cm prediction, averaged across the three Bayesian architectures, evaluated without **(e)** and with **(f)** Supervised-PCA dimensionality reduction. **g, h** Spearman correlation coefficients for $dG_{H2O}$ prediction, averaged across the three Bayesian architectures, evaluated without **(g)** and with **(h)** Supervised-PCA dimensionality reduction.

### S4.3 SARS-COV-2 Receptor Binding Domain

We further evaluated the Bayesian architectures using the SARS-CoV-2 RBD[24] benchmark, which contains multiple experimentally measured properties and therefore provides an independent assessment of multitask learning performance (Supplementary Figure **7**). Consistent with the results obtained for both RhlA and Thermonuclease, BLL generally achieved the highest predictive performance prior to dimensionality reduction, while the differences between architectures became considerably smaller after applying SPCA. As a result, all three Bayesian approaches exhibited similar predictive accuracy across the evaluated properties.

The analysis of numerical representations showed that the best-performing embedding depended on the prediction task before dimensionality reduction, reflecting the heterogeneous characteristics of the dataset (Supplementary Figure **7**c,e,g). After SPCA, performance differences between representations decreased substantially, and multiple reduced representations achieved similarly high predictive accuracy(Supplementary Figure **7**d,f,h). This indicates that the reduced feature space captured the information required for accurate prediction while diminishing the influence of the initial embedding choice.

Taken together, the SARS-CoV-2 RBD benchmark confirms the trends observed across the other case studies. Although the optimal numerical representation remained dataset-dependent, SPCA consistently produced more stable predictive performance across both Bayesian architectures and protein representations. The reproducibility of these observations across three independent multitask datasets suggests that the benefits of dimensionality reduction are not specific to a single benchmark but represent a general characteristic of the evaluated protein engineering prediction tasks.

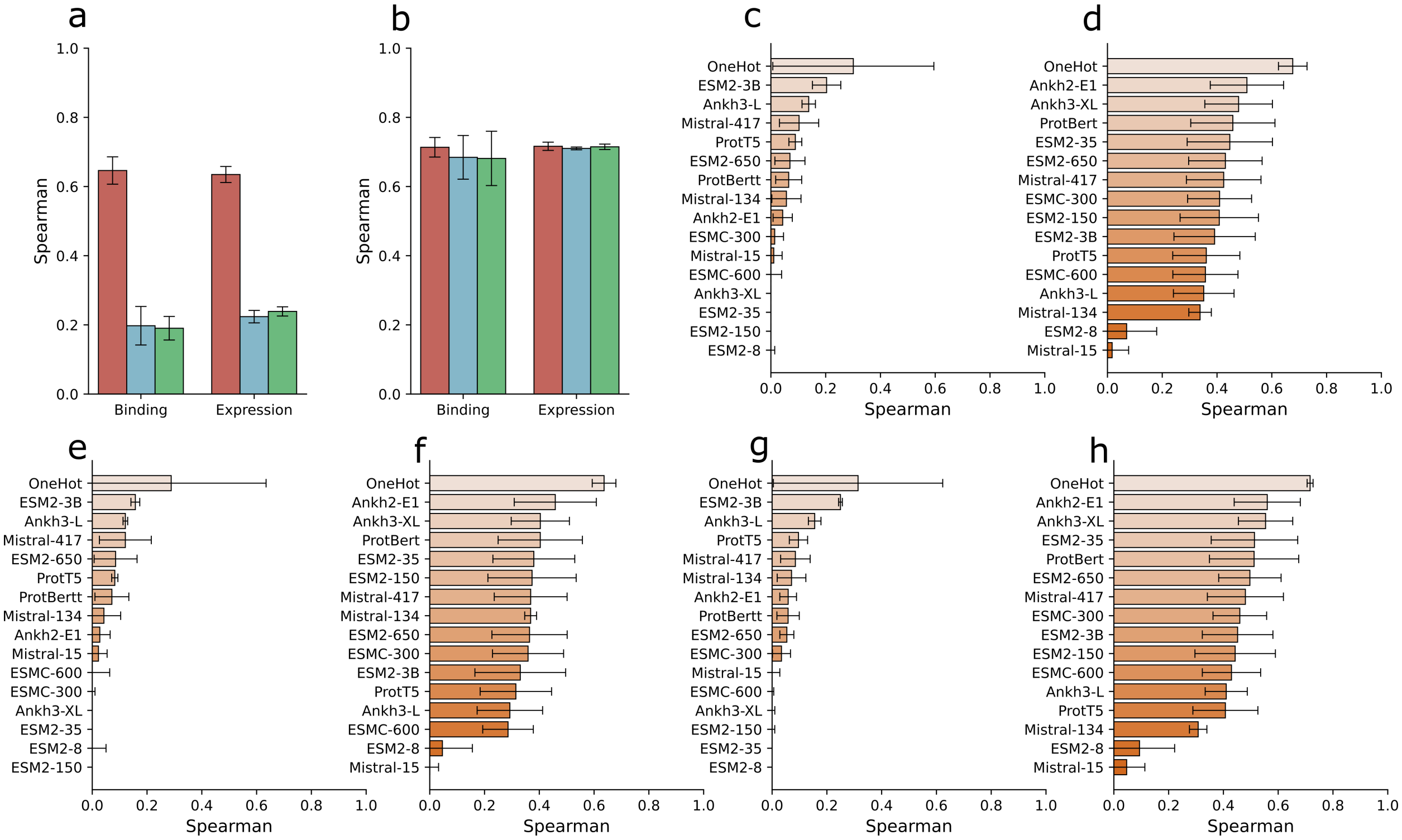

**Supplementary Figure 7**: **Predictive performance on the SARS-COV-2 Receptor Binding Domain benchmark. a, b** Spearman correlation coefficients obtained by the three best-performing numerical representations for Bayesian Last Layer (BLL, red), Monte Carlo Dropout (MCD, blue), and Hierarchical Bayesian Neural Networks (HBN, green) for the joint prediction of Binding and Expression, evaluated without **(a)** and with **(b)** Supervised-PCA dimensionality reduction. **c, d** Average Spearman correlation coefficients across the three Bayesian architectures for the combined Binding and Expression prediction, evaluated without **(c)** and with **(d)** Supervised-PCA dimensionality reduction. **e, f** Spearman correlation coefficients for Binding prediction, averaged across the three Bayesian architectures, evaluated without **(e)** and with **(f)** Supervised-PCA dimensionality reduction. **g, h** Spearman correlation coefficients for Expression prediction, averaged across the three Bayesian architectures, evaluated without **(g)** and with **(h)** Supervised-PCA dimensionality reduction.